\documentclass[11pt,a4paper]{article}
\usepackage{xcolor}
\usepackage{amsmath}
\usepackage{amssymb}
\usepackage{appendix}
\usepackage{multirow}
\usepackage{geometry}
\usepackage{float}
\usepackage[export]{adjustbox}
\usepackage{cite}

\usepackage{hyperref}

\usepackage{graphicx}
\usepackage[normalem]{ulem}

\newcommand{\Ceff}[1]{{\cal C}^{\rm eff}_{#1}}

\newcommand{\Cc}[1]{{\cal C}_{#1}}

\newcommand{\op}{\mathcal{O}}
\newcommand{\GeV}{\,{\rm GeV}}
\newcommand{\heff}{\mathcal{H}_{\rm eff}}

\begin{document}

\begin{flushright}
\today
\end{flushright}

\vspace{0.3cm}

\begin{center}
{\bf {\Large On the impact of meson mixing
on  $B_s\to\phi ee$ angular \\[0.2em] observables at low $q^2$}}

\vspace{0.5cm}

S\'ebastien Descotes-Genon, Ioannis Plakias, Olcyr Sumensari

\vspace{0.2cm}

\emph{Universit\'e Paris-Saclay, CNRS/IN2P3, IJCLab, 91405 Orsay, France}

\end{center}

\vspace{0.5cm}

\begin{abstract}
Decays based on the $b\to s\gamma$ transition are expected to yield left-handed photons in the Standard Model, but could be particularly sensitive to New Physics contributions modifying the short-distance Wilson coefficients $\Cc{7}$ and $\Cc{7'}$ defined in the low-energy Effective Field Theory. These coefficients can be determined by combining observables of several modes, among which $B_s\to\phi ee$ at low $q^2$, in a kinematic range where the photon-pole is dominant. We investigate the impact of $B_s-\bar{B}_s$ mixing on the angular observables available for this mode, which induces a time-dependent modulation governed by interference terms between mixing and decay that are sensitive to the moduli and phases of $\Cc{7}$ and $\Cc{7'}$. These interference terms can be extracted through a time-dependent analysis but they also affect time-integrated observables. In particular, we show that the asymmetries $A_T^{(2)}$ and $A_T^{(\text{Im});CP}$ receive terms of potentially similar size from mixing-independent and mixing-induced terms, and we discuss how the constraints coming from the angular analysis of $B_s\to\phi ee$ at low $q^2$ should be interpreted in the presence of mixing.
\end{abstract}

\section{Introduction}

Processes mediated by flavor-changing neutral currents are loop-suppressed in the Standard Model (SM), making them useful probes of New Physics. A prominent example is the $b\to s\mu\mu$ transition, for which several discrepancies have been observed in the past several years.
Indeed, LHCb data exhibits deviations close to $3\sigma$ from the SM expectation in the $P^\prime_5$ angular observable of $B \to K^*\mu\mu$ decay~\cite{Descotes-Genon:2013vna}, and tensions are also seen in branching ratios of $b\to s\mu\mu$ exclusive decays~\cite{LHCb:2013ghj,LHCb:2013tgx,LHCb:2014cxe,LHCb:2020dof,LHCb:2020lmf}. Deviations are also hinted at in Belle data for $B\to K^*\mu\mu$~\cite{Belle:2016fev}.
Global fits~\cite{Alguero:2021anc, Altmannshofer:2021qrr, Hurth:2021nsi, Geng:2021nhg, Ciuchini:2020gvn} indicate that these deviations can be described consistently in the Low-Energy Effective Field Theory (LEFT) at the $\mu=m_b$ scale, if we assume that New Physics (NP) modifies the short-distance contributions of the leading SM operators and/or their chirality-flipped versions. Several NP scenarios can provide an equally good description of the data, with the interesting possibility of a $V-A$ structure of the NP contribution, which would be similar to the SM one.

The closely related $b\to s\gamma$ transition is also known to be a powerful probe of NP effects~\cite{Atwood:1997zr}. This flavour-changing neutral current is also suppressed in the SM and can be affected by NP effects at the loop level. Although the number of observables is limited for $b\to s\gamma$, its theoretical description is much simpler than its $b\to s\ell\ell$ counterpart. The $V-A$ structure of the weak interaction in the SM means that the photon will be  dominantly produced with a left-handed polarisation, and that the right-handed polarisation is highly suppressed by the ratio of quark masses $m_s/m_b$~\cite{Atwood:1997zr}. This is encoded at the level of the LEFT by the suppression of the Wilson coefficient $\Cc{7'}$ compared to the leading one $\Cc{7}$. However, NP in right-handed currents can alter the SM hierarchy among photon polarisations, which would provide a very interesting hint on the nature of the NP responsible for the deviations observed in $b\to s\ell\ell$ processes. These NP effects could generate additional phases leading to new forms of CP-violation in $b\to s\gamma$ decays. Finally, in the context of global explanations of $b\to s\ell\ell$ and $b\to c\tau\nu$ anomalies within an EFT framework, an interesting NP scenario consists in ``large'' contributions to $b\to c\tau\nu$ and $b\to s\tau\tau$ operators~(see e.g.~Ref.~\cite{Alonso:2015sja,Capdevila:2017iqn,Buttazzo:2017ixm,Angelescu:2018tyl,Angelescu:2021lln,Cornella:2021sby}), which could feed into NP lepton-flavour universal contributions to $b\to s\ell\ell$~\cite{Crivellin:2018yvo,Cornella:2021sby}, but also into possible NP contributions to $b\to s\gamma$ through radiative corrections.

Many different approaches have been proposed to extract information on the photon polarisation in $b\to s\gamma$, and more generally on the values of $\Cc{7}$ and $\Cc{7'}$~\cite{Paul:2016urs}.
The branching ratio for the inclusive $B\to X_s\gamma$ decay has reached a high level of precision both theoretically and experimentally~\cite{Misiak:2015xwa}. Time-dependent analysis of $B\to K^*\gamma$ and $B_s\to\phi\gamma$ can be performed to extract mixing-induced CP-asymmetries~\cite{Belle:2006pxp,BaBar:2008okc,LHCb:2016oeh} containing relevant information on $\Cc{7}$ and $\Cc{7'}$~\cite{Atwood:2004jj,Grinstein:2005nu,Ball:2006cva,Muheim:2008vu}. The baryonic mode $\Lambda_b\to\Lambda\gamma$ also provides interesting constraints~\cite{Hiller:2001zj,Legger:2006cq,LHCb:2021byf}. Other approaches such as converted photons~\cite{Grossman:2000rk} and asymmetries in $B\to K_1(\to K\pi\pi)\gamma$~\cite{Gronau:2001ng,Gronau:2002rz,Kou:2010kn,Gronau:2017kyq,Akar:2018zhv} have also been considered.

Another possibility is to consider semileptonic decays based on the $b\to see$ transition at very low-$q^2$ values, in a kinematic range that is not reachable for the muonic channel. In this regime, the photon pole dominates and the transverse polarisations provide leading contributions to all observables. Therefore, the angular analysis can provide information on the photon polarisation through the available observables, namely transverse asymmetries~\cite{Kruger:2005ep,Becirevic:2011bp}. The main advantage of these observables is that they are rather theoretically clean, since the relevant form factors cancel out completely as $q^2$ approaches the photon pole. LHCb has performed very accurate measurements of these asymmetries for $B\to K^\ast ee$ with 9 fb$^{-1}$, providing significant constraints on $\Cc{7}$ and $\Cc{7'}$~\cite{LHCb:2020dof}. In particular, these results provide the leading constraints on the real and imaginary parts of $C_{7^\prime}$ as of today. A similar analysis is certainly possible for $B_s\to\phi ee$, given that angular analyses are available for both $B\to K^\ast$ and $B_s\to\phi$ modes in the case of the $b\to s\mu\mu$ transition~\cite{LHCb:2020lmf,LHCb:2021xxq}.

Interestingly, the $B_s\to \phi\ell\ell$ decays are not self-tagging and their final states ($K^+K^-$ or $K_SK_L$) are CP-eigenstates. Therefore, $B_s-\bar{B}_s$ mixing must be included to fully describe these decays, which may provide non-negligible corrections given the size of mixing parameters in the $B_s$ system. The impact of mixing has been discussed for various $B$-meson decays~\cite{Descotes-Genon:2007iri,Bobeth:2008ij,DeBruyn:2012wk,Descotes-Genon:2022gcp}.
For $B\to V\ell\ell$~\cite{Descotes-Genon:2015hea} and $B\to P\ell\ell$~\cite{Descotes-Genon:2020tnz} modes, the time-dependence of the branching ratio will involve new observables that depend on the interference between mixing and decay. This may provide additional information about the relative moduli and phases of the relevant transversity amplitudes. The modifications induced by neutral-meson mixing for the $B_s\to\phi\ell\ell$ mode have already been considered in Refs.~\cite{Bobeth:2008ij,Descotes-Genon:2015hea}. However, the hierarchy of amplitudes change at low-$q^2$ values due to the dominance of the photon pole and the impact of $B_s$-meson mixing on all the accessible observables is not necessarily intuitive. This article is thus focused on assessing the impact of $B_s$-mixing on the low-$q^2$ observables for $B_s\to\phi ee$ and demonstrating that additional information can be gathered from the interference between mixing and decay occurring in this mode.

The remainder of this article is organized as follows. In Sec.~\ref{sec:nomixing}, we discuss the angular analysis of $B_s\to\phi ee$, firstly neglecting neutral-meson mixing, and we determine the observables of interest close to the photon pole. In Sec.~\ref{sec:mixing}, we deal with mixing and consider the time-dependent analysis of $B_s\to\phi ee$ and how time-integrated angular observables will also carry information on the relative size and phase of $\Cc{7}$ and $\Cc{7'}$. In Sec.~\ref{sec:numerical}, we perform a thorough numerical analysis, checking the range of validity of the photon-pole approximation, the constraints obtained from the low-$q^2$ angular observables and their sensitivity to NP effects, showing the importance of taking into account $B_s-\bar{B}_s$ mixing, before concluding in Sec.~\ref{sec:conclusion}. Appendices are devoted to conventions regarding kinematics and helicity amplitudes, to collect lengthy expressions of the angular coefficients in the presence of mixing, as well as to discuss further transverse asymmetries which prove more difficult to reach experimentally.

\section{$B_s\to \phi ee$ in the absence of mixing}\label{sec:nomixing}

\subsection{Low-energy effective theory Hamiltonian}

The $b\to s\ell\ell$ transitions are described by the usual LEFT Hamiltonian with SM operators, in addition to the NP ones with a chirality-flipped, scalar or tensor structure~\cite{Buchalla:1995vs,Altmannshofer:2008dz}:
\begin{equation}
{\cal H}_{\rm eff} = -\frac{4 G_F}{\sqrt{2}}
\bigg[
\lambda_u\,[\Cc1 (\op_1^c-\op_1^u) +
\Cc2 (\op_2^c-\op_2^u)]
+\lambda_t\sum_{i\in I} \Cc{i} \op_{i}
\bigg] +\mathrm{h.c.}\ ,
\end{equation}
where $\lambda_q = V_{qb} V_{qs}^\ast$ and $I \in \{1c,2c, 3,4,5,6, 8, 7^{(\prime)},9^{(')}\ell,10^{(\prime)}\ell,S^{(\prime)}\ell,P^{(\prime)}\ell,T^{(\prime)}\ell \}$.
In the following, we neglect doubly Cabibbo suppressed contributions, of relative size of $\mathcal{O}(\lambda^2)\simeq 4\%$ where $\lambda$ is the usual parameter of the Wolfenstein parametrisation of the Cabibbo-Kobayashi-Maskawa (CKM) matrix. This leads us to neglect the contributions proportional to $\lambda_u$  in $\heff$.
The operators $\op_{1,..,6}$ and $\op_8$ are hadronic operators of the type
$(\bar s \Gamma b)(\bar q \Gamma' q)$ and
$(\bar s \gamma^{\mu\nu} T_a P_R b) G_{\mu\nu}^a$, respectively. These operators are not likely to receive very large contributions from NP, as they would appear in non-leptonic $B$ decay amplitudes~\footnote{See Refs.~\cite{Brod:2014bfa,Lenz:2019lvd,Jager:2019bgk} for a discussion of low-energy constraints on these operators.}.
The main operators of interest
$\op_{7^{(\prime)},9^{(\prime)},10^{(\prime)}}$ are then given by
\begin{align}\label{effops}
\begin{split}
{\cal O}_{9^{(\prime)}\ell}  &=  \frac{e^2}{(4\pi)^2} [\bar{s} \gamma^\mu P_{L(R)} b] [\bar{\ell} \gamma_\mu \ell] \,,\qquad\quad
{\cal O}_{10^{(\prime)}\ell}  =  \frac{e^2}{(4\pi)^2} [\bar{s} \gamma^\mu P_{L(R)} b] [\bar{\ell} \gamma_\mu \gamma_5 \ell]\,,\\[0.3em]
{\cal O}_{7^{(\prime)}} &= \frac{e}{(4\pi)^2} m_b [\bar{s} \sigma^{\mu\nu} P_{R(L)} b] F_{\mu\nu}\,.
\end{split}
\end{align}
In the SM, and at a scale $\mu_b=\op(m_b)$, the Wilson coefficients of interest
Eq.~(\ref{effops}) are $\Cc7^\text{SM}(\mu_b)\simeq -0.3$, $\Cc{7'}^\text{SM}(\mu_b)\simeq -0.006$ (suppressed by $m_s/m_b$ compared to $\Cc7$),
$\Cc9^\text{SM}(\mu_b)\simeq 4.1$ and $\Cc{10}^\text{SM}(\mu_b)\simeq -4.3$, which are identical for $\ell=e$ and $\ell=\mu$ due to the universality of the SM gauge-couplings to leptons. Note, in particular, that these Wilson coefficients might be affected by complex NP contributions, which can also violate LFU, being different for $\ell=e$ and $\ell=\mu$, see e.g.~Ref.~\cite{Becirevic:2020ssj}.

When considering actual matrix elements describing $b\to s\ell\ell$ or $b\to s\gamma$ transitions, certain combinations
of these Wilson coefficients naturally arise. One repeatedly encounters the regularisation-scheme independent combinations of Wilson coefficients 
$\Ceff{j=7,8}=\Cc{j}+\sum_{i=1}^6 y_{ji} \Cc{i}$, where $y_i$ are pure numbers given by RGE~\cite{Buchalla:1995vs}. Moreover, one has to take into account 
long-distance contributions coming from four-quark operators and corresponding to charm-loop contributions. As we will discuss below, these contributions can be absorbed into $q^2$- and final-state-dependent ``effective" Wilson-coefficients, corresponding to a vector coupling to leptons. 
This long-distance part is thus absorbed in $\Cc{7}$ for the $b\to s\gamma$ transition, while the customary choice for $b\to s\ell\ell$ transitions is $\Cc{9}$ (whereas $\Cc{7}$ gets only redefined by ultraviolet contributions required by renormalization), see e.g. Ref.~\cite{Altmannshofer:2008dz}.

\subsection{$B_s\to \phi(\to K^+ K^-)\ell\ell$ angular coefficients}\label{sec:angularnomix}

In the following, we consider the $B_s\to \phi(\to K^+ K^-)\ell\ell$ decays, which has already been considered by LHCb in the $\ell=\mu$ channel~\cite{LHCb:2021xxq} \footnote{In principle, the $B_s\to \phi(\to K_S K_L)\ell\ell$ mode could also be considered, but it is far more challenging experimentally.}. 
In the absence of mixing, we can give the general expression for the angular
coefficients using the general formalism of Refs.~\cite{Kruger:2005ep,Altmannshofer:2008dz}, with the angular convention specified in Appendix~\ref{app:conventions},
\begin{eqnarray}\label{eq:dist}
\frac{d^4\Gamma(B_s\to \phi (\to K^+ K^-)\ell\ell)}{dq^2\,d\!\cos\theta_K\,d\!\cos\theta_l\,d\phi}&=&\frac9{32\pi} \bigg[
J_{1s} \sin^2\theta_K + J_{1c} \cos^2\theta_K + J_{2s} \sin^2\theta_K\cos 2\theta_l\nonumber\\[0.2em]
&&\hspace{-4.2cm} + J_{2c} \cos^2\theta_K \cos 2\theta_l
+ J_3 \sin^2\theta_K \sin^2\theta_l \cos 2\phi + J_4 \sin 2\theta_K \sin 2\theta_l \cos\phi\nonumber\\[0.2em]
&&\hspace{-4.2cm}+ J_5 \sin 2\theta_K \sin\theta_l \cos\phi 
+ J_{6s} \sin^2\theta_K\cos\theta_l + {J_{6c} \cos^2\theta_K}  \cos\theta_l\\[0.2em]  
&&\hspace{-4.2cm}+ J_7 \sin 2\theta_K \sin\theta_l \sin\phi  + J_8 \sin 2\theta_K \sin 2\theta_l \sin\phi 
+ J_9 \sin^2\theta_K \sin^2\theta_l \sin 2\phi \bigg]\,,\nonumber
\end{eqnarray}
which depend on the invariant mass of the lepton pair ($q^2$), and three angles
that we denote $\theta_\ell$, $\theta_K$ and $\phi$. The angle $\theta_\ell$ is defined between the lepton $\ell^+$ direction with respect to the opposite of the direction of flight of the $B_s$ meson in the $\ell^+\ell^-$ centre-of-mass frame. Similarly, $\theta_K$ is defined as the angle of the $K^-$ meson with respect to the opposite direction of flight of the $B_s$-meson in the $\phi$-meson rest frame, see Fig.~\ref{fig:angular} in Appendix~\ref{app:conventions}. Note that this choice of kinematics differs from the one considered in the LHCb analysis for self-tagging decays~\cite{LHCb:2015ycz,LHCb:2020dof,Gratrex:2015hna,Becirevic:2016zri} (where $\theta_\ell$ would be associated with $\ell^-$ for $B_q$ decays and $\ell^+$ for $\bar{B}_q$ decays), and from the theory convention~\cite{Altmannshofer:2008dz} (where this angle would be associated to $\ell^-$ for both $B_q$ and $\bar{B}_q$ decays), but it is the most appropriate when we discuss non-self-tagging modes such as $B_s\to\phi\ell\ell$~\cite{Descotes-Genon:2015hea}.

The coefficients of the distribution $J_i(q^2)$ contain interference terms
of the form ${\rm Re}[A_XA_Y^*]$ and ${\rm Im}[A_XA_Y^*]$ between the eight
transversity amplitudes defined in Appendix~\ref{app:transversity},
\begin{equation}
\big{\lbrace} A_{0}^{L},\ A_{0}^{R},\ A_{||}^{L},\ A_{||}^{R},\ A_{\perp}^{L},\ A_{\perp}^{R},\ A_{t},\ A_S\big{\rbrace} \,,
\end{equation}
which are given by
\begin{eqnarray}
J_{1s}  & = & \frac{(2+\beta_\ell^2)}{4} \left[|A_{\perp}^{L}|^2 + |A_{||}^{L}|^2 +|A_{\perp}^{R}|^2 + |A_{||}^{R}|^2 \right]
+ \frac{4 m_\ell^2}{s} {\rm Re}\left(A_{\perp}^{L}A_{\perp}^{R*} + A_{||}^{L}A_{||}^{R*}\right)\,,\nonumber\\[0.25em]
J_{1c}  & = &  |A_{0}^{L}|^2 +|A_{0}^{R}|^2  + \frac{4m_\ell^2}{s} \left[|A_t|^2 + 2{\rm Re}(A_{0}^{L}A_{0}^{R*}) \right] + \beta_\ell^2\, |A_S|^2 \,,\nonumber\\[0.25em]
J_{2s} & = & \frac{ \beta_\ell^2}{4}\left[ |A_{\perp}^{L}|^2+ |A_{||}^{L}|^2 + |A_{\perp}^{R}|^2+ |A_{||}^{R}|^2\right],
\hspace{0.92cm}    J_{2c}  = - \beta_\ell^2\left[|A_{0}^{L}|^2 + |A_{0}^{R}|^2 \right]\,,\nonumber\\[0.25em]
J_3 & = & \frac{1}{2}\beta_\ell^2\left[ |A_{\perp}^{L}|^2 - |A_{||}^{L}|^2  + |A_{\perp}^{R}|^2 - |A_{||}^{R}|^2\right],
\qquad   J_4  = \frac{1}{\sqrt{2}}\beta_\ell^2\left[{\rm Re} (A_{0}^{L}A_{||}^{L*} + A_{0}^{R}A_{||}^{R*} )\right],\nonumber \\[0.25em]
J_5 & = & \sqrt{2}\beta_\ell\,\Big[{\rm Re}(A_{0}^{L}A_{\perp}^{L*} - A_{0}^{R}A_{\perp}^{R*} ) - \frac{m_\ell}{\sqrt{s}}\,
{\rm Re}(A_{||}^{L} A_S^*+ A_{||}^{R*} A_S) \Big]\,,\nonumber\\[0.25em]
J_{6s} & = &  2\beta_\ell\left[{\rm Re} (A_{||}^{L}A_{\perp}^{L*} - A_{||}^{R}A_{\perp}^{R*}) \right]\,,
\hspace{2.25cm} J_{6c} = 4\beta_\ell\, \frac{m_\ell}{\sqrt{s}}\, {\rm Re} (A_{0}^{L} A_S^*+ A_{0}^{R*} A_S)\,,\nonumber\\[0.25em]
J_7 & = & \sqrt{2} \beta_\ell\, \Big[{\rm Im} (A_{0}^{L}A_{||}^{L*} - A_{0}^{R}A_{||}^{R*} ) +
\frac{m_\ell}{\sqrt{s}}\, {\rm Im} (A_{\perp}^{L} A_S^* - A_{\perp}^{R*} A_S)) \Big]\,,\nonumber\\[0.25em]
J_8 & = & \frac{1}{\sqrt{2}}\beta_\ell^2\left[{\rm Im}(A_{0}^{L}A_{\perp}^{L*} + A_{0}^{R}A_{\perp}^{R*})\right]\,,
\hspace{1.9cm} J_9 = \beta_\ell^2\left[{\rm Im} (A_{||}^{L*}A_{\perp}^{L} + A_{||}^{R*}A_{\perp}^{R})\right] \,,
\label{Js}
\end{eqnarray}
where $\beta_\ell=\sqrt{1- 4m_\ell^2/q^2}$.
In the massless lepton limit, we have $J_{1s}=3J_{2s}$ and $J_{1c}=-J_{2c}$ and $J_{6c}=0$. The expression for the various transversity amplitudes and Wilson coefficients can be found e.g.~in Ref.~\cite{Altmannshofer:2008dz}. 

Similar expressions hold for the CP-conjugate decay $\bar{B}_s\to \bar{\phi}(\to K^+ K^-) \ell \ell$,
with angular coefficients $\bar{J}_i$ involving amplitudes denoted by $\bar A_X$, and obtained from the $A_X$
by conjugating all CP-odd ``weak'' phases~\footnote{This is opposite to the notation used in Ref.~\cite{Altmannshofer:2008dz} for $B$
and $\bar{B}$ decays, but in agreement with general discussions on CP-violation and the discussions of refs.~\cite{Descotes-Genon:2015hea,Descotes-Genon:2020tnz}.}.
These weak phases appear in the CKM matrix elements involved in the normalisation of the weak effective Hamiltonian as well as the short-distance Wilson coefficients $C_i$ present in the definition of the Wilson coefficients. These amplitudes will involve in general~\cite{Altmannshofer:2008dz}
\begin{equation}
\bar{A}[V_{tb}V_{ts}^*,{\cal C}_i,F_j,h_j]\,, \qquad A[V_{tb}^*V_{ts},{\cal C}_i^*,F_j,h_j]\,,
\end{equation}
where $F_i$ are local form-factors describing $\langle V|\op_{i}|B\rangle$ and $h_j$ are non-local contributions from charm loops~\footnote{We assume that there are no significant complex NP contributions to the short-distance four-quark Wilson coefficients $\Cc{1\ldots 6}$ so that these contributions carry no weak phases.}.

The form of the angular distribution
for the CP-conjugated decay depends on the way the kinematical variables
are defined. In the case in which the \emph{same} conventions are used for the lepton angle irrespective of whether the decaying meson is a $B_s$ or a ${\bar B}_s$, we have~\cite{Altmannshofer:2008dz,Descotes-Genon:2015hea}
\begin{eqnarray}
\frac{d\Gamma[B_s\to \phi(\to K^+ K^-) \ell^+\ell^-]}
{dq^2\  d\!\cos\theta_\ell\ d\!\cos\theta_M\ d\phi}&=&
\sum_i J_i(q^2) f_i(\theta_\ell,\theta_M,\phi)\,,
\label{Gamma}\\
\frac{d\Gamma[\bar{B}_s\to \phi(\to K^+ K^-) \ell^+\ell^-]}
{dq^2\  d\!\cos\theta_\ell\ d\!\cos\theta_M\ d\phi}&=&
\sum_i \zeta_i\bar{J}_i(q^2) f_i(\theta_\ell,\theta_M,\phi)
=\sum_i \tilde{J}_i(q^2) f_i(\theta_\ell,\theta_M,\phi)\,,
\label{Gammabar}
\end{eqnarray}
where $f_i(\theta_\ell,\theta_M,\phi)$ are defined by Eq.~(\ref{eq:dist}).
These expressions feature two different angular coefficients $\widetilde J_i$ and $\bar J_i$ which are CP conjugates of $J_i$:

\begin{itemize}
\item The angular coefficients $\widetilde{J}_i$ formed by replacing $A_X$ by $\widetilde A_X\equiv A_X(\bar{B}_s\to f_{CP})$
(without CP-conjugation applied on $f_{CP}$), which will appear in the study of time evolution due to mixing,
where both $B_s$ and $\bar B_s$ decay into the same final state $f_{CP}$.

\item The angular coefficients $\bar{J}_i$, obtained by considering $\bar A_X\equiv A_X(\bar B_s \to \overline f_{CP})$ (with CP-conjugation applied to $f_{CP}$), which can be obtained from $A_X$ by changing the sign of all weak phases, and arise naturally when discussing CP violation (and CP conjugation) from the theoretical point of view.

\end{itemize}
As discussed in Ref.~\cite{Descotes-Genon:2015hea,Descotes-Genon:2020tnz},
we have $\widetilde A_X = \eta_X \bar A_X$, with $\eta_X$ given by
\begin{equation}
\eta_X=\eta \quad \text{for}\quad  X=L0,L||,R0,R||,t \quad; \quad \eta_X=-\eta \quad \text{for}\quad X=L\!\perp,R\!\perp,S\ ,
\label{etaX}
\end{equation}
with $\eta=1$ in the $B_s\to \phi(\to K^+ K^-)\ell\ell$ case. 
Therefore $\bar{J}_i$ can be obtained from $J_i$ by changing the sign of all weak phases, and $\tilde{J}_i=\zeta_i \bar{J}_i$ with
\begin{equation}\label{eq:zetadef}
\zeta_i=1\quad{\rm for}\quad i=1s,1c,2s,2c,3,4,7\ ; \qquad
\zeta_i=-1\quad{\rm for}\quad i=5,6s,6c,8,9\ .
\end{equation}

Since the final state is not self-tagging, an untagged measurement of the differential decay rate
(e.g. at LHCb, where the production asymmetry is tiny) yields
\begin{equation}
\frac{d\Gamma(B_s\to f_{CP})+d\Gamma(\bar{B}_s\to f_{CP})}
{dq^2\  d\!\cos\theta_\ell\ d\!\cos\theta_K\ d\phi}
=\sum_i [J_i+\widetilde{J}_i]  f_i(\theta_\ell,\theta_K,\phi)
=\sum_i [J_i+\zeta_i \bar{J}_i]  f_i(\theta_\ell,\theta_K,\phi)\ ,
\label{G+Gb}
\end{equation}
which involves CP-averages for some of the observables, but CP-asymmetries for the others.
The difference between the two decay rates  can only be measured through flavour-tagging, and it
involves $J_i - \widetilde J_i = J_i-\zeta_i \bar{J}_i$, providing access to other averages and asymmetries,
\begin{equation}
\frac{d\Gamma(B_s\to f_{CP})-d\Gamma(\bar{B}_s\to f_{CP})}
{dq^2\  d\!\cos\theta_\ell\ d\!\cos\theta_K\ d\phi}
=\sum_i [J_i-\widetilde{J}_i]  f_i(\theta_\ell,\theta_K,\phi)
=\sum_i [J_i-\zeta_i \bar{J}_i]  f_i(\theta_\ell,\theta_K,\phi)\ .
\label{GmGb}
\end{equation}
We emphasize that we have neglected any effect coming from mixing at this stage and that we assumed that there was no asymmetry in the production of $B_s$ and $\bar{B}_s$ meson.

\subsection{$B_s\to \phi (\to K^+ K^-) ee$ angular observables}\label{sec:bstophieeangular}

We focus now on the $B_s\to \phi ee$ observables that can be extracted experimentally. To this purpose, we define~\cite{Altmannshofer:2008dz}
\begin{equation}
S_i=(J_i+\bar{J}_i)\Big/\frac{d(\Gamma+\bar\Gamma)}{dq^2}\,, \qquad A_i=(J_i-\bar{J}_i)\Big/\frac{d(\Gamma+\bar\Gamma)}{dq^2}\,, 
\end{equation}
where we stress again that $J$ and $\bar{J}$ would be associated with $B_s$ and $\bar{B}_s$ decays, respectively, which is the opposite  
of  the notation of Ref.~\cite{Altmannshofer:2008dz}. The binned version of these expressions would correspond to integrating the numerator and the denominator over the bin in $q^2$ separately before taking the ratio~\cite{Descotes-Genon:2012isb,Descotes-Genon:2013vna}. We have:
\begin{eqnarray}\label{eq:DefObs}
\frac{1}{d(\Gamma+\overline{\Gamma})/dq^2}\frac{d^3(\Gamma+\overline{\Gamma})}{d\cos\theta_\ell\, d\cos\theta_K\, d\phi}
&=&
\frac9{32\pi} \bigg[
S_{1s} \sin^2\theta_K + S_{1c} \cos^2\theta_K + S_{2s} \sin^2\theta_K\cos 2\theta_l\nonumber\\[0.25em]
&&\hspace{-4.2cm} + S_{2c} \cos^2\theta_K \cos 2\theta_l
+ S_3 \sin^2\theta_K \sin^2\theta_l \cos 2\phi + S_4 \sin 2\theta_K \sin 2\theta_l \cos\phi\nonumber\\[0.25em]
&&\hspace{-4.2cm}+ A_5 \sin 2\theta_K \sin\theta_l \cos\phi 
+ A_{6s} \sin^2\theta_K\cos\theta_l +  {A_{6c} \cos^2\theta_K}  \cos\theta_l\\[0.25em]
&&\hspace{-4.2cm}+ S_7 \sin 2\theta_K \sin\theta_l \sin\phi  + A_8 \sin 2\theta_K \sin 2\theta_l \sin\phi 
+ A_9 \sin^2\theta_K \sin^2\theta_l \sin 2\phi \bigg]\,,\nonumber
\end{eqnarray}

\noindent The expression of the differential decay rate yields
\begin{equation}
\frac{d\Gamma}{dq^2}=\frac{3}{4}(2J_{1s}+J_{1c})-\frac{1}{4}(2J_{2s}+J_{2c})\,,
\end{equation}
leading to the relation
\begin{equation}
\frac{3}{4}(2S_{1s}+S_{1c})-\frac{1}{4}(2S_{2s}+S_{2c})=1
\end{equation}

\noindent We can rewrite this angular description in a way closer to the expression in the massless limit:
\begin{eqnarray}
\frac{1}{d(\Gamma+\overline{\Gamma})/dq^2}\frac{d^3(\Gamma+\overline{\Gamma})}{d\cos\theta_\ell\, d\cos\theta_K\, d\phi}
&=&
\frac9{32\pi} \bigg[
\left[\frac{3}{4}F_T+\Delta_{1s} \right]\sin^2\theta_K
+ \left[F_L + \Delta_{1c}\right] \cos^2\theta_K \nonumber\\
&&\hspace{-4.2cm} +\frac{1}{4}F_T\sin^2\theta_K\cos 2\theta_l - F_L \cos^2\theta_K \cos 2\theta_l\nonumber\\
&&\hspace{-4.2cm}
+\frac{1}{2}F_T A_{T}^{(2)} \sin^2\theta_K \sin^2\theta_l \cos 2\phi + S_4 \sin 2\theta_K \sin 2\theta_l \cos\phi\nonumber\\
&&\hspace{-4.2cm}+ A_5 \sin 2\theta_K \sin\theta_l \cos\phi 
+ F_T A_T^{(\text{Re});CP} \sin^2\theta_K\cos\theta_l +  {A_{6c} \cos^2\theta_K}  \cos\theta_l\\
&&\hspace{-4.2cm}+ S_7 \sin 2\theta_K \sin\theta_l \sin\phi  + A_8 \sin 2\theta_K \sin 2\theta_l \sin\phi 
+\frac{1}{2}F_T A_T^{(\text{Im});CP}\sin^2\theta_K \sin^2\theta_l \sin 2\phi \bigg]\,,\nonumber
\end{eqnarray}
where we define the following observables~\cite{Altmannshofer:2008dz},
\begin{equation}
F_L=-S_{2c}\,, \qquad F_T=4 S_{2s} \,,
\end{equation}
and the three transverse asymmetries involving only $A_\perp$ and $A_{||}$ amplitudes which will prove dominant near the photon pole, by analogy with the $B\to K^*ee$ case~\cite{Kruger:2005ep,Becirevic:2011bp},
\begin{equation}\label{eq:defasymmetries}
 A_{T}^{(2)}=\frac{S_{3}}{2S_{2s}}\,, \qquad A_T^{(\text{Re});CP}=\frac{A_{6s}}{4S_{2s}}\,, \qquad
 A_{T}^{(\text{Im});CP}=\frac{A_{9}}{2S_{2s}}\,,
\end{equation}
where we explicitly introduce a $CP$ superscript to indicate that we build the quantity from a $CP$-asymmetry, and
we introduce the corrections vanishing in the massless limit,
\begin{equation}
\Delta_{1s}=S_{1s}-3S_{2c}\,,\quad\qquad \Delta_{1c}=S_{1c}+S_{2c}\,.
\end{equation}
Defined in this way, the numerator of both $F_L$ and $F_T$ involves only the expected transversity amplitudes, but the denominator involves not only the sum over the modulus square of all amplitudes, but also $\mathcal{O}(m_\ell^2)$-suppressed interference terms as well as scalar contributions coming from $J_{1s}$ and $J_{1c}$. Therefore, we obtain $F_L+F_T=1+\mathcal{O}(m_\ell^2,(C_S-C_S')^2)$. 

In the massless lepton limit, we have
\begin{eqnarray}\label{eq:AT2ampl}
A_{T}^{(2)}&=&\frac{|A^L_\perp|^2+|A^R_\perp|^2-|A_{||}^L|^2-|A_{||}^R|^2+(A \leftrightarrow \bar{A})}{|A^L_\perp|^2+|A^R_\perp|^2+|A_{||}^L|^2+|A_{||}^R|^2+(A \leftrightarrow \bar{A})}\,,\\[0.45em]
A_T^{(\text{Re});CP}&=&\frac{2\,{\rm Re}\big{[}(A_{||}^LA_\perp^{L*}-A_{||}^R A_\perp^{R*})-(A \leftrightarrow \bar{A})\big{]}}{|A^L_\perp|^2+|A^R_\perp|^2+|A_{||}^L|^2+|A_{||}^R|^2+(A \leftrightarrow \bar{A})}\,,\\[0.45em]
A_T^{(\text{Im});CP}&=&-\frac{2\,{\rm Im}\big{[}(A_{||}^LA_\perp^{L*}+A_{||}^R A_\perp^{R*})-(A \leftrightarrow \bar{A})\big{]}}{|A^L_\perp|^2+|A^R_\perp|^2+|A_{||}^L|^2+|A_{||}^R|^2+(A \leftrightarrow \bar{A})}\,.
\label{eq:ATImCPampl}
\end{eqnarray}

\noindent where $(A\leftrightarrow \bar{A})$ corresponds to the replacement of $\smash{A_i^{L(R)}}$ by their CP-conjugate counterparts $\smash{\bar{A}_i^{L(R)}}$. It proves interesting to rewrite the various transversity amplitudes according to the Wilson coefficients involved ($X=\perp,||$)~\cite{Altmannshofer:2008dz},
\begin{equation}\label{eq:ampldecomposition}
A^X_{L,R}=e^{i\beta_s}\Big[\rho^X_{79}e^{i\phi^X_{79}}+h^X e^{i\delta^X}\mp \rho^X_{10}e^{i\phi^X_{10}}\Big]\,, \qquad    
\bar{A}^X_{L,R}=e^{-i\beta_s}\Big[\rho^X_{79}e^{-i\phi^X_{79}}+h^X e^{i\delta^X}\mp \rho^X_{10}e^{-i\phi^X_{10}}\Big]\,,
\end{equation}
showing the contribution proportional to $\Cc{7^{(')},9^{(')}}$, the function $h$ stemming from $c\bar{c}$ loops and the contribution proportional to $\Cc{10^{(')}}$, respectively.
In this decomposition, $\rho^X$ denote moduli, and $\delta^X$ and $\phi^X$ denote CP-even ``strong'' and ``weak" phases, respectively, whereas $\beta_s$ is an overall phase coming from the CKM factors $\lambda_t$ in the LEFT Hamiltonian. We obtain then
\begin{eqnarray}
A_{T}^{(2)}&=&
\frac{1}{D}\Big[
(\rho^\perp_{79})^2+(\rho^\perp_{10})^2+(h^\perp)^2+2h^\perp\rho^\perp_{79}\cos\delta^\perp \cos\phi^\perp_{79} - (\perp \leftrightarrow || )\Big]\,,\nonumber\\[0.4em] 
A_T^{(\text{Re});CP}&=&
-\frac{2}{D}\Big[
h^\perp\rho^{||}_{10}\sin\delta^\perp\sin\phi^{||}_{10}
+(\perp \leftrightarrow || )
\Big]\,,
\\[0.4em]
A_T^{(\text{Im});CP}&=&\frac{1}{D}\Big[ \rho^\perp_{79}\rho^{||}_{79} \sin(\phi^{\perp}_{79}-\phi^{||}_{79}) +\rho^\perp_{10}\rho^{||}_{10} \sin(\phi^{\perp}_{10}-\phi^{||}_{10})-
2 h^\perp\rho^{||}_{79}\cos\delta^\perp\sin\phi^{||}_{79}- (\perp \leftrightarrow || )\Big{]}\,,\nonumber
\end{eqnarray}
with
\begin{eqnarray}
\label{eq:Dparameter}
 D&=&(\rho^\perp_{79})^2+(\rho^\perp_{10})^2+(h^\perp)^2+2\rho^\perp_{79}h^\perp\cos \delta^\perp \cos\phi^\perp_{79} +(\perp \leftrightarrow || )\,.
\end{eqnarray}
We see that $A_T^{(\text{Re});CP}$ is non-vanishing only in the presence of a complex contribution to $\Cc{10(')}$, and its interpretation requires accurate knowledge of the strong phase generated by the $c\bar{c}$ loops. Therefore, this observable will not be particularly useful to constrain $\Cc{7}$ and $\Cc{7'}$ at low $q^2$, as will become clear in the following. On the contrary, the two other asymmetries contain terms involving only $\Cc{7(')}$ and will thus be relevant at low $q^2$, as already established in the $B\to K^\ast ee$ case~\cite{Becirevic:2011bp}.

Finally, let us mention that the definitions of the three transverse asymmetries agree with the ones in Ref.~\cite{Becirevic:2011bp} derived for the self-tagging modes $\bar{B}\to \bar{K}^{*0}ee$. We stress that the observables that can be accessed here are more limited than in self-tagging modes where both CP-averaged quantities and CP-asymmetries can be determined in principle.
In particular, for $B_s\to \phi ee$, if we measure only $d\Gamma+d\bar\Gamma$, we have only  $A_T^{(\text{Re});CP}$ and $A_T^{(\text{Im});CP}$ and we cannot access the CP-averaged quantities $A_T^{(\text{Re})}$ and $A_T^{(\text{Im})}$ that could be measured in $B\to K^*ee$~\cite{Becirevic:2011bp,LHCb:2015ycz,LHCb:2020dof}:
\begin{equation}\label{eq:defasymmbis}
 A_T^{(\text{Re})}=\frac{S_{6s}}{4S_{2s}}\,, \qquad
 A_{T}^{(\text{Im})}=\frac{S_{9}}{2S_{2s}}\,.
\end{equation}
In the case of $B_s\to \phi ee$, these asymmetries could be determined only if we measure $d\Gamma-d\bar\Gamma$ (thus requiring flavour tagging). It would also provide access to the asymmetry
\begin{equation}\label{eq:defasymmbis2}
    A_{T}^{(2);CP}=\frac{A_{3}}{2S_{2s}}\,.
\end{equation}
We discuss these last three observables in more detail in Appendix~\ref{app:asymmbis} \footnote{For completeness, we stress that LHCb measurements of $B\to K^*ee$ at low $q^2$~\cite{LHCb:2020dof}
were performed with kinematic conventions leading to 
$A_T^{(2)}$, $A_T^{(\text{Re})}$ and $A_T^{(\text{Im});CP}$ (denoted $A_T^{(\text{Im})}$ in Ref.~\cite{LHCb:2020dof}), in addition to the longitudinal polarisation $F_L$ and the branching ratio.}.

\subsection{Photon-pole approximation for $B_s\to\phi ee$ in the absence of mixing}\label{sec:Bsnomixing}

We focus now on the $B_s\to\phi ee$ decays at low $q^2$, where the photon pole is expected to dominate. This means considering $q^2$ as negligible compared to $m_\phi^2$ and $m_{B_s}^2$, but of the same order compared to $4\,m_e^2$. Before providing the complete expressions for the $B_s\to\phi ee$ observables in the presence of mixing, we provide a rough estimate of the various angular observables near the photon pole. This exercise will allow us to identify which quantities are potentially sensitive to $C_{7^{(\prime)}}$ in this regime. To this purpose, we introduce a spurious quantity $\epsilon\simeq 0.05$ that we will use to determine which contributions are large, i.e.~$\mathcal{O}(\epsilon^0)$, and which ones are suppressed, i.e.~$\mathcal{O}(\epsilon^{n\geq 1})$.
If we consider the typical range for low-$q^2$ analysis at LHCb~\cite{LHCb:2020dof}, we notice that in the SM
\begin{equation}
\label{eq:epsilon-def}
\frac{4m_e^2}{q^2}\sim \epsilon^0 \,,\quad \frac{m_\phi^2}{m_B^2}\sim \epsilon\,,\quad
 \frac{q^2}{m_\phi^2}\sim \epsilon\,, \quad
\frac{\Cc{7^{(')}}}{\Cc{9^{(')}}}\sim \epsilon\,, \quad \frac{\Cc{9^{(')}}}{\Cc{10^{(')}}}\sim \epsilon^0\,,
\end{equation}
where the last two relations assume that NP does not change the hierarchy of Wilson coefficients within the SM.
Experimentally, $q^2$ is currently much larger than the lower end of the kinematic range, and we could have used an alternative counting such as $4m_e^2/q^2\sim \epsilon^2$ for all practical purposes, but we prefer to consider the possibility that measurements at even lower $q^2$ can be performed in the future~\footnote{For reference, the $q^2$-interval considered in the latest $B\to K^* ee$ LHCb analysis is $[0.0008,0.257]~\mathrm{GeV}^2$~\cite{LHCb:2020dof}.}.
We further notice that we assume that $q^2$ is much smaller than the masses of the hadrons involved, but the assumption that $q^2/m_\phi^2=\mathcal{O}(\epsilon)$ does not hold when we get closer to upper range of the current analyses, as $q^2$ should not be significantly larger than
$0.05~\GeV^2$. Nonetheless, this will have no impact either on the qualitative discussions derived below or on the binned observables whose integrals over $q^2$ are dominated by the photon-pole region. Keeping the leading term in this expansion will correspond to what is called the photon-pole approximation in the following, and the accuracy of this approximation will be numerically assessed in Sec.~\ref{sec:validityphotonpole}.

Up to correction of order $\mathcal{O}(\epsilon)$ and using the results of Ref.~\cite{Altmannshofer:2008dz}, we have for both $L$ and $R$ amplitudes,
\begin{equation}
   \bar{A}_{\perp}= -N_0\frac{m_b}{\sqrt{q^2}} e^{-i\beta_s}\bar{C}_+T_1(0) [1+\mathcal{O}(\epsilon)]\,, \qquad
    \bar{A}_{\parallel} = N_0\frac{m_b}{\sqrt{q^2}} e^{-i\beta_s}\bar{C}_-T_1(0)[1+\mathcal{O}(\epsilon)] \,,
\end{equation}
leading to
\begin{eqnarray}
 \tilde{A}_{\perp}= N_0\frac{m_b}{\sqrt{q^2}} e^{-i\beta_s}\bar{C}_+T_1(0) [1+\mathcal{O}(\epsilon)] \,, &\quad& A_{\perp} =-N_0\frac{m_b}{\sqrt{q^2}} e^{i\beta_s}C_+ T_1(0)[1+\mathcal{O}(\epsilon)] \,,\\[0.25em]
    \tilde{A}_{\parallel} = N_0\frac{m_b}{\sqrt{q^2}} e^{-i\beta_s}\bar{C}_-T_1(0)[1+\mathcal{O}(\epsilon)] \,, &\quad& A_{\parallel} = N_0\frac{m_b}{\sqrt{q^2}} e^{i\beta_s}C_-T_1(0)[1+\mathcal{O}(\epsilon)]  \,.
\end{eqnarray}
where $\tilde{A}_{\perp}$ and $\tilde{A}_{\parallel}$ are defined similarly to Ref.~\cite{Descotes-Genon:2015hea,Descotes-Genon:2020tnz}. The effective Wilson coefficients are
\begin{equation}
C_{\pm} \equiv (\Cc{7}^{\text{eff}} \pm \Cc{7'}^{\text{eff}})^* + \frac{h_\pm(0)}{T_1(0)}\,,\qquad\quad
\bar{C}_{\pm} \equiv \Cc{7}^{\text{eff}} \pm \Cc{7'}^{\text{eff}} + \frac{h_\pm(0)}{T_1(0)}\,,
\end{equation}
where $h_\pm(q^2)$ are non-local contributions coming from $c\bar{c}$ loops contributions, assuming once again
that there are no complex NP contributions to four-quark Wilson coefficients $\Cc{i=1\ldots 6}$.
The contribution from $c\bar{c}$ pairs is estimated to be small, amounting to
less than $10\,\%$ of the SM value according to the computations from Refs.~\cite{Ball:2006eu,Muheim:2008vu,Khodjamirian:2010vf}.

We have introduced the angle $\beta_s=-\arg(-V_{cs} V^*_{cb}/(V_{ts} V^*_{tb}))$ so that $V_{ts}=-|V_{ts}|e^{i\beta_s}$.
The normalization
\begin{equation}
N_0 = \frac{|V_{tb} V_{ts}|G_F\alpha_{\mathrm{em}}(m_{B_s}^2-m_{\phi}^2)^{3/2}}{2^{7/2}\sqrt{3}\pi^{5/2}m_{B_s}^{3/2}}
\end{equation}
is independent of $q^2$ and, in the following, we will also use the $q^2$-dependent quantity
\begin{equation}
N' = N_0\frac{m_b}{\sqrt{q^2}}T_1(0)\,.
\end{equation}
We see that the transverse amplitudes $A_\perp$ and $A_{\parallel}$ count as $\mathcal{O}(\epsilon^{-1})$ due to the photon pole whereas the other amplitudes read
 \begin{equation}
 A_0 =\mathcal{O}(\epsilon^0)\,,  \qquad
 A_t =\mathcal{O}(\epsilon^0)\,, \qquad
 A_S=\mathcal{O}(\epsilon)\,, 
 \end{equation}
where we assume that the contribution from pseudoscalar operators to $A_t$ is at most of the same order as that of $\Cc{10}$, corresponding to $m_b(\Cc{P}-\Cc{P'})/\Cc{10}=\mathcal{O}(\epsilon^0)$.

As expected, the amplitudes for transverse polarisations become dominant near the photon pole since the longitudinal polarisations, as well as the other contributions, are forbidden when the photon becomes real. According to this counting, the leading mixing-independent angular coefficients in Ref.~\cite{Altmannshofer:2008dz} and in Eq.~(\ref{eq:dist}) are $\mathcal{O}(\epsilon^{-2})$,
\begin{alignat}{3}
    J_{1s} &\sim \frac{3}{2}\alpha_{-1/3} |N'|^2  (|C_+|^2+|C_-|^2)\,,  \qquad 
    & J_{2s} &\sim \frac{1}{2}\alpha_{1}  |N'|^2 (|C_+|^2+|C_-|^2)\,,\\[0.25em]
    J_3 &\sim \alpha_{1} |N'|^2  (|C_+|^2-|C_-|^2)\,, \qquad
    & J_9 &\sim -2\alpha_{1} |N'|^2\text{Im}[C_-^*C_+]  \,, \nonumber
\end{alignat}
where we denote
\begin{equation}
\alpha_n=1-n\frac{4m_e^2}{q^2}\,.
\end{equation}
The other angular coefficients are subleading: 
\begin{equation}
J_5\sim J_7\sim J_8 = \mathcal{O}(\epsilon^{-1})\,,
\qquad\quad
J_{1c} \sim J_{2c} \sim J_4 \sim J_{6s} \sim J_{6c}=\mathcal{O}(\epsilon^0)\,.
\end{equation}
The CP-conjugates $\bar{J}_i$ are obtained by replacing $C_\pm$ by $\bar{C}_\pm$.
This means that the non-vanishing observables in the photon-pole approximation are $S_{2s}$, $S_3$, $A_9$ (from $d\Gamma+d\bar\Gamma$) and $A_{2s}$, $A_3$, $S_9$ (from $d\Gamma-d\bar\Gamma$), confirming that 
$A_T^{(2)}$ and $A_T^{(\text{Im});CP}$ are asymmetries of primary interest at low $q^2$, which will be studied in full generality in Sec.~\ref{sec:numerical}.

Lastly, a comment is in order concerning $A_T^{(\text{Re});CP}$, which vanishes in the limit $q^2\to 0$, but could still carry some relevant information on NP away from the photon pole. In Sec.~\ref{sec:bstophieeangular}, we demonstrated that this quantity would be sensitive to the interference between weak phases in $\Cc{7^{(')},9^{(')}}$ and $\Cc{10^{(')}}$. This scenario lies outside of the scope of the NP scenarios considered here (i.e.,~NP only in $\Cc{7^{(')}}$), but it could obviously be of interest and worth measuring to probe more general NP scenarios involving CP-violating phases in other Wilson coefficients.

\section{$B_s\to \phi ee$ in the presence of mixing}\label{sec:mixing}

\subsection{Time-dependent angular analysis}

Once we consider neutral-meson mixing into their antiparticles and decaying into CP eigenstates, we can use Refs.~\cite{Descotes-Genon:2015hea,Descotes-Genon:2020tnz} to determine the structure of the time-dependent angular analysis of $B_s\to \phi ee$. The induced time dependence 
can be expressed in terms of the two different angular coefficients $\widetilde J_i$ and $\bar J_i$. The time-dependent angular coefficients are obtained by replacing the time-independent amplitudes with the time-dependent
ones in the definition of the $J_i$ coefficients
\begin{equation}
J_i(t) = J_i \big(A_X\to A_X(t)\big)\ ,\qquad
\widetilde J_i(t) = J_i \big(A_X\to \widetilde A_X(t)\big)\ .
\label{subst}
\end{equation}
We consider the combinations $J_i(t) \pm \widetilde J_i(t)$ appearing in the sum and difference of time-dependent
decay rates in Eqs.~(\ref{G+Gb})-(\ref{GmGb}),
\begin{eqnarray}\label{eq:J+Jt}
J_i(t)+\widetilde J_i(t) &=&e^{-\Gamma t}\Big[(J_i + \widetilde J_i)\cosh(y\Gamma t) - h_i \sinh(y\Gamma t)\Big]\ ,\\[2mm]
J_i(t)-\widetilde J_i(t) &=&e^{-\Gamma t}\Big[(J_i - \widetilde J_i)\cos(x\Gamma t) - s_i \sin(x\Gamma t)\Big]\ ,
\label{eq:J-Jt}
\end{eqnarray}
where $x\equiv \Delta m_s/\Gamma_s$ and $y\equiv \Delta \Gamma_s/(2\Gamma_s)$, and we have defined a new set of angular
coefficients $s_i,h_i$ related to the time-dependent angular distribution. Their expressions can be found in the Appendix of Ref.~\cite{Descotes-Genon:2015hea} and correspond to interference terms between two decay amplitudes and the mixing phase $q/p=\exp(i\phi)$. For simplicity, and in agreement with the current measurements~\cite{Lenz:2012az,Charles:2015gya,Charles:2020dfl}, we assume that $B_s-\bar{B}_s$ mixing receives no significant NP contributions and thus $\phi=2\beta_s$, leading to a cancellation of the (small) $\beta_s$ phases between decay amplitudes and mixing in the interference terms $h$ and $s$~\footnote{Note that one can easily extend our expressions to include a NP phase entering $B_s$ mixing from the expressions given   in Ref.~\cite{Descotes-Genon:2015hea}}.

One can also consider measurements integrated over time. As discussed in Ref.~\cite{Descotes-Genon:2015hea}, if we consider set-ups such as $B$-factories where the $B_{s}$-meson production occurs coherently through $\Upsilon(5S)$ decays, the effect of mixing is washed out after time integration. On the other hand, at LHCb, and more generally, at machines producing pairs of $B$ hadrons in an incoherent manner such as LHC or FCCee, we have the time-integrated expressions,
\begin{eqnarray}
\label{eq:<J+Jt>Had}
\langle J_i + \widetilde J_i\rangle_{\rm inc}
 &=& \frac{1}{\Gamma} \left[\frac{1}{1-y^2} \times(J_i+\widetilde J_i)-\frac{y}{1-y^2}\times h_i\right]\ ,\\
\label{eq:<J-Jt>Had}
\langle J_i - \widetilde J_i\rangle_{\rm inc}
 &=& \frac{1}{\Gamma}\Bigg[\frac{1}{1+x^2} \times (J_i-\widetilde J_i) - \frac{x}{1+x^2} \times s_i \Bigg]\ .
\end{eqnarray}

\noindent For $B_s$ mesons, we have $x= 26.81\pm 0.08$ and $y=0.0675\pm 0.004$~\cite{Workman:2022ynf}, and the effect is immediately present for $B_s\to \phi \ell\ell$ since $\phi$ decays into a $CP$ eigenstate of two kaons ($K^+K^-$ or $K_SK_L$). The effect is in general small for $J_i+\tilde{J}_i$ (i.e.~of $\mathcal{O}(y)$) unless the value of the observable without mixing is suppressed (which can be the case for some of the observables at low $q^2$). The effect is expected to be much larger for $J_i-\tilde{J}_i$ (i.e.~of $\mathcal{O}(1)$).

\subsection{Time-dependence of the observables in the photon-pole approximation}\label{sec:timedepobs}

Similarly to Sec.~\ref{sec:Bsnomixing}, we first consider the photon-pole approximation, expanding the various angular coefficients to the leading order in the spurious coefficient $\epsilon$, see Eq.~\eqref{eq:epsilon-def}. This exercise will allow us to quantify the bulk of the mixing effects near the photon pole, which will be refined in the numerical analysis of Sec.~\ref{sec:numerical}. The mixing-dependent part of the angular observables is given by the photon-pole approximation of the expressions in Ref.~\cite{Descotes-Genon:2015hea},
\begin{eqnarray}
\begin{split}    
    s_{1s} &\sim -3\alpha_{-1/3}|N'|^2\,\text{Im}[\bar{C}_+C_+^*-\bar{C}_-C_-^*] \,, & \qquad
   s_{2s} & \sim -\alpha_{1} |N'|^2\,\text{Im}[\bar{C}_+C_+^*-\bar{C}_-C_-^*]\,,\\[0.45em]
    s_3 &\sim -2\alpha_{1} |N'|^2\,\text{Im}[\bar{C}_+C_+^*+\bar{C}_-C_-^*]\,, &
    s_9 & \sim -2\alpha_{1} |N'|^2 \,\text{Re}[\bar{C}_-C_+^*+\bar{C}_+^*C_-] \,,
\end{split}
\end{eqnarray}
and
\begin{eqnarray}
\begin{split}    
    h_{1s} &\sim -3\alpha_{-1/3}|N'|^2\,\text{Re}[\bar{C}_+C_+^*-\bar{C}_-C_-^*] \,, & \qquad
    h_{2s} & \sim -\alpha_{1} |N'|^2\,\text{Re}[\bar{C}_+C_+^*-\bar{C}_-C_-^*]\,,\\[0.45em]
    h_3 & \sim -2\alpha_{1} |N'|^2\,\text{Re}[\bar{C}_+C_+^*+\bar{C}_-C_-^*]\,, & 
    h_9 & \sim 2\alpha_{1} |N'|^2 \,\text{Im}[\bar{C}_-C_+^*-\bar{C}_+^*C_-] \,,
\end{split}
\end{eqnarray}
whereas  the other obervables $s_i$ and $h_i$ with $i\in \lbrace 1c,2c,4,5,6s,6c,7,8\rbrace$ 
are subleading, with the same power of $\epsilon$ as the corresponding angular coefficients $J$. These observables may provide additional information on $C_+$ and $C_-$, and in particular their weak phases. 

In principle, the coefficients listed above can be obtained by considering a flavour-tagged time-dependent angular analysis of $B_s\to\phi ee$, for instance, at the Belle II experiment at the $\Upsilon(5S)$ peak producing coherent $B_s$--$\bar{B_s}$ pairs~\cite{Belle-II:2018jsg}. Such a measurement would allow us to assess the coefficients of Eqs.~(\ref{eq:J+Jt})-(\ref{eq:J-Jt}) in terms of the time difference $t$ between the flavour-tagging decay on one side and the decay into the final state of interest ($\phi ee$) on the other side.
As an illustration, for the $B_s\to \phi ee$ branching fraction, we find that
\begin{eqnarray}
\frac{d(\Gamma+\bar\Gamma)}{dq^2}
	&&=e^{-\Gamma t}\left[\frac{1}{2}[3(J_{1s}+\tilde{J}_{1s})-(J_{2s}+\tilde{J}_{2s})]\cosh(y\Gamma t)
	   -\frac{1}{2}[3h_{1s}-h_{2s}]\sinh(y\Gamma t) \right]\,,
\end{eqnarray}
yielding, after expanding in $\epsilon$,
\begin{eqnarray}
\frac{d(\Gamma+\bar\Gamma)}{dq^2}&& \sim  e^{-\Gamma t}\alpha_{-1/2}  |N'|^2 \Big{[}2\left(|C_+|^2+|C_-|^2+|\bar{C}_+|^2+|\bar{C}_-|^2\right)\cosh(y\Gamma t)\\[0.2em]
	   &&\hspace{7.5em}+4\text{Re}[\bar{C}_+C_+^*-\bar{C}_-C_-^*] \sinh(y\Gamma t) \Big{]}\,,\nonumber 
\end{eqnarray}
which incidentally shows that the asymmetry ${\cal A}_{\Delta \Gamma}(B_s\to \phi\gamma)$ measured by LHCb~\cite{LHCb:2016oeh} is closely related to the $q^2\to 0$ limit of $3h_{1s}-h_{2s}$. Similarly, $S_{K^*\gamma}$ measured at the $B$-factories~\cite{Belle:2006pxp,BaBar:2008okc} would be related to the $q^2\to 0$ limit of the $B\to K^*\gamma$ equivalent of $3s_{1s}-s_{2s}$. 

Finally, we emphasize that the determination of the other coefficients $s$ and $h$ would require a time-dependent angular analysis which might prove challenging unless a very large sample of $B_s\to\phi ee$ decays has been gathered. We will not focus on this type of determination in the following, but the reader can easily identify  the role played by $h$ and $s$ in the expressions that we will provide afterwards.

\subsection{Time-integrated observables in the photon-pole approximation}\label{sec:timeintobs}

For incoherent production (for instance at LHCb, but also $Z$-factories like FCCee or CEPC), it proves interesting to consider the time-integrated observables $S_i$ and $A_i$, defined by integrating over time the numerator and the denominator defining each ratio, i.e.
\begin{equation}
X=\frac{N}{D}\,, \qquad \mathrm{with}\qquad \langle X\rangle_{\rm inc}=\frac{\int_0^\infty dt\ N}{\int_0^\infty dt\ D}\,.
\end{equation}
As discussed in Refs.~\cite{Descotes-Genon:2015hea,Descotes-Genon:2020tnz},  the same integral would range from $-\infty$ to $\infty$ in the case of a coherent production, washing away mixing-induced terms. This does not occur in the case of an incoherent production, allowing one to probe mixing-induced terms through time-integrated measurements. Due to the $\mathcal{O}(\epsilon^{-2})$ singular behaviour of the branching ratio in the denominator defining $S_i$ and $A_i$,
\begin{equation}
\small \int_0^\infty dt\ \frac{d(\Gamma+\bar\Gamma)}{dq^2}\sim
 \frac{1}{\Gamma} \alpha_{-1/2} |N'|^2 \frac{2}{1-y^2} \left[[|C_+|^2+|C_-|^2+|\bar{C}_+|^2+|\bar{C}_-|^2]
	   +y\cdot 2\text{Re}[\bar{C}_+C_+^*-\bar{C}_-C_-^*]\right] \,,
\end{equation}
the observables corresponding associated with subleading numerators $\mathcal{O}(\epsilon^{-1},\epsilon^0)$ turn out to vanish in the photon-pole approximation
\begin{equation}
    \langle S_i \rangle_\mathrm{inc} \to 0\quad\text{and}\quad  \langle A_i \rangle_\mathrm{inc} \to 0\qquad \text{for}\quad i\in \lbrace 1c,2c,4,5,6s,6c,7,8\rbrace\,,
\end{equation}
whereas the non-vanishing observables are $S_{1s}$, $S_{2s}$, $S_3$ and $S_9$, as well as $A_{1s}$, $A_{2s}$, $A_3$ and $A_9$, with the corresponding expressions collected in Appendix~\ref{app:photon-pole}~\footnote{We recall that  $S_{1s},S_{2s},S_3,A_9$ are accessible through an untagged measurement at LHCb, but that is not possible for $A_{1s},A_{2s},A_3,S_9$.}. By using these expressions, we can show that the considered transverse asymmetries at low $q^2$ will be 

\begin{eqnarray}
     \langle A_T^{(2)}\rangle_{\rm inc} &\sim& 
     \frac{[|C_+|^2+|\bar{C}_+|^2 - |C_-|^2- |\bar{C}_-|^2]+y\cdot 2 \text{Re}[\bar{C}_+C_+^*+\bar{C}_-C_-^*]}
    {[|C_+|^2 +|\bar{C}_+|^2 + |C_-|^2 +|\bar{C}_-|^2 ]+ y\cdot 2 \text{Re}[{\bar{C}_+C_+^*-\bar{C}_-C_-^*]}}\,,\\[0.45em]
     \langle A_T^{(\text{Im});CP}\rangle_{\rm inc}  &\sim & 
      2\frac{-\text{Im}[C_-^* C_+ - \bar{C}_-^* \bar{C}_+]- y\cdot  \text{Im}[\bar{C}_-C_+^*-\bar{C}_+^*C_-]}
    {[|C_+|^2 +|\bar{C}_+|^2 + |C_-|^2 +|\bar{C}_-|^2 ]+ y\cdot 2 \text{Re}[{\bar{C}_+C_+^*-\bar{C}_-C_-^*]}}  \,,\\[0.45em]
     \langle A_T^{(\text{Re});CP}\rangle_{\rm inc}  &\sim & 0 \,,
\end{eqnarray}

\noindent whereas $\langle F_L\rangle_{\rm inc}\to 0$ and $\langle F_T\rangle_{\rm inc}\to 1$, as expected since the photon becomes real and only its transverse polarisations should be relevant (this also holds before integrating the transverse and longitudinal polarisations over time). In particular, we find that the angular observables
$A_T^{(2)}$ and $\smash{A_T^{(\text{Im});CP}}$ feature interesting sensitivities to $C_+$ and $C_-$ through mixing, which are in principle different from the constraints obtained from other observables. 

\subsection{Constraints on $\Cc{7}$ and $\Cc{7'}$ near the SM point}\label{sec:nearSM}

We can derive further simplified expressions in the vicinity of the SM. We will assume that we can neglect the $c\bar{c}$ contributions $h_\pm(0)$ and that the NP corrections to $\Cc{7^{(')}}$ may carry weak phases but do not modify the SM hierarchy (i.e.~we do not consider scenarios with very large right-handed currents). We can then expand in $\eta=\mathcal{O}(\epsilon)$, counting
$y\sim 1/x\sim \Cc{7'}/\Cc{7} =\mathcal{O}(\eta)$ and keep only $\mathcal{O}(\eta)$ contributions. We obtain:
\begin{eqnarray}
    \langle S_{1s}\rangle_{\rm inc} &\sim& \frac{3}{4} \frac{\alpha_{-1/3}}{\alpha_{-1/2}}\,,  \qquad\qquad
    \langle S_{2s}\rangle_{\rm inc} \sim \frac{1}{4}\frac{\alpha_{1}}{\alpha_{-1/2}} \,,\\[0.35em] 
  \langle S_3\rangle_{\rm inc} &\sim&\frac{\alpha_{1}}{\alpha_{-1/2}}
   \frac{1}{(\text{Re}[\Cc{7}])^2+(\text{Im}[\Cc{7}])^2} \\
   &&\qquad\times
   \left[\text{Re}[\Cc{7}]\text{Re}[\Cc{7'}]+\text{Im}[\Cc{7}]\text{Im}[\Cc{7'}]
    +\frac{y}{2}[(\text{Re}[\Cc{7}])^2-(\text{Im}[\Cc{7}])^2]\right]\,,\nonumber\\[0.35em]
   \langle  S_9\rangle_{\rm inc} &\sim& \frac{1}{2} \frac{\alpha_{1}}{\alpha_{-1/2}}\frac{1}{x}\frac{(\text{Re}[\Cc{7}])^2-(\text{Im}[\Cc{7}])^2}{(\text{Re}[\Cc{7}])^2+(\text{Im}[\Cc{7}])^2}\,,
   \end{eqnarray}
and
\begin{eqnarray}
   \langle A_{1s}\rangle_{\rm inc} &\sim&  \langle A_{2s}\rangle_{\rm inc}  \sim 0\,,\\[0.35em] 
    \langle  A_3\rangle_{\rm inc} &\sim& \frac{\alpha_{1}}{\alpha_{-1/2}}
     \frac{1}{x}\frac{\text{Re}[\Cc{7}]\text{Im}[\Cc{7}]}{(\text{Re}[\Cc{7}])^2+(\text{Im}[\Cc{7}])^2}\,,\\[0.35em] 
 \langle A_9\rangle_{\rm inc} &\sim& \frac{\alpha_{1}}{\alpha_{-1/2}} \frac{\text{Re}[\Cc{7}]\text{Im}[\Cc{7'}]-\text{Re}[\Cc{7'}]\text{Im}[\Cc{7}]  
 -y\text{Re}[\Cc{7}] \text{Im}[\Cc{7}] }{(\text{Re}[\Cc{7}])^2+(\text{Im}[\Cc{7}])^2} \,.
\end{eqnarray}
Therefore, we find that
\begin{eqnarray}\label{eq:AT2nearSM}
\langle A_T^{(2)}\rangle_{\rm inc}&\sim&   \frac{2}{(\text{Re}[\Cc{7}])^2+(\text{Im}[\Cc{7}])^2} \\
   &&\qquad\times
   \left[\text{Re}[\Cc{7}]\text{Re}[\Cc{7'}]+\text{Im}[\Cc{7}]\text{Im}[\Cc{7'}]
    +\frac{y}{2}[(\text{Re}[\Cc{7}])^2-(\text{Im}[\Cc{7}])^2]\right]\,,\nonumber\\[0.35em]
\langle A_T^{(\text{Re});CP}\rangle_{\rm inc}   &\sim & 0\,,\\[0.35em]
\label{eq:ATimCPnearSM}
\langle A_T^{(\text{Im});CP}\rangle_{\rm inc}
&\sim&
\frac{2}{(\text{Re}[\Cc{7}])^2+(\text{Im}[\Cc{7}])^2} \\
   &&\qquad\times
\Big{[}\text{Re}[\Cc{7}]\text{Im}[\Cc{7'}]-\text{Re}[\Cc{7'}]\text{Im}[\Cc{7}]  -y\text{Re}[\Cc{7}] \text{Im}[\Cc{7}]\Big{]} \,.\nonumber 
\end{eqnarray}

\noindent Within this limited framework, we already see that 
\begin{itemize}
\item Mixing effects do not impact $ \langle S_{1c}\rangle_{\rm inc}$, $ \langle S_{2c}\rangle_{\rm inc}$, $ \langle A_{1c}\rangle_{\rm inc}$, $ \langle A_{2c}\rangle_{\rm inc}$, which are linked to $s$ and $h$ coefficients probed directly by the time-evolution of the $B_s\to\phi\gamma$ decay rate.
\item The contributions induced via mixing compete with the mixing-independent  part for $\langle S_3\rangle_{\rm inc}$ and
$\langle A_9\rangle_{\rm inc}$, or equivalently  $\smash{\langle A_T^{(2)}\rangle_{\rm inc}}$ and 
$\smash{\langle A_T^{(\text{Im});CP}\rangle_{\rm inc}}$. Therefore, one should be careful in the interpretation of $\smash{\langle A_T^{(2)}\rangle_{\rm inc}}$ and $\smash{\langle A_T^{(\text{Im});CP}\rangle_{\rm inc}}$ to consider the non-negligible mixing-induced effects to obtain the correct constraints on $\Cc{7}$ and $\Cc{7'}$. In particular, this effect is already important for $\langle A_T^{(2)}\rangle_{\rm inc}$ in the SM.
\item Mixing effects become the dominant contributions (suppressed by $1/x$) in $\langle A_3\rangle_{\rm inc}$ and $ \langle  S_9\rangle_{\rm inc}$, or equivalently $\smash{\langle A_T^{(2);CP}\rangle_{\rm inc}}$ and $\smash{\langle A_T^{(\text{Im})}\rangle_{\rm inc}}$, as discussed in Appendix~\ref{app:asymmbis}. Their deviations from zero would provide information on the imaginary part of $\Cc{7}$ and $\Cc{7'}$ that is complementary to other quantities. More specifically, comparing Eqs.~(\ref{eq:AT2CPnearSM})-(\ref{eq:ATimnearSM}) in Appendix~\ref{app:asymmbis} and
Eqs.~(\ref{eq:AT2nearSM})-(\ref{eq:ATimCPnearSM}), we see that they coincide with the mixing-induced part of $\langle A_T^{(2)}\rangle_{\rm inc}$ and $\langle A_T^{(\text{Im});CP}\rangle_{\rm inc}$ in this limit.
\end{itemize}
The difference between $A_T^{(2)}$  for $B_s\to \phi ee$ and $B\to K^*ee$ 
(and similarly $A_T^{(\text{Im});CP}$)
can provide additional constraints on $\Cc{7}$,
\begin{align}
\begin{split}
\Delta_T^{(2)}&\equiv\langle A_T^{(2)}\rangle_{\rm inc}(B_s\to\phi ee)
-\langle A_T^{(2)}\rangle_{\rm inc}(B\to K^*ee)\\[0.25em]
&=y\dfrac{(\text{Re}[\Cc{7}])^2-(\text{Im}[\Cc{7}])^2}{(\text{Re}[\Cc{7}])^2+(\text{Im}[\Cc{7}])^2}\,,
\end{split}
\end{align}
and
\begin{align}
\begin{split}\Delta_T^{(\text{Im});CP}&\equiv\langle A_T^{(\text{Im});CP}\rangle_{\rm inc}(B_s\to\phi ee)
-\langle A_T^{(\text{Im});CP}\rangle_{\rm inc}(B\to K^*ee)\\[0.25em]
&=-2y\frac{\text{Re}[\Cc{7}]\text{Im}[\Cc{7}]}{(\text{Re}[\Cc{7}])^2+(\text{Im}[\Cc{7}])^2}\,,
\end{split}
\end{align}
which is also given by
$\smash{\langle A_T^{(\text{Im})}\rangle_{\rm inc}}$ and $\smash{\langle A_T^{(2);CP}\rangle_{\rm inc}}$, respectively, up to a suppression by a factor $1/x$ rather than $y$.

Clearly, the above interpretation changes if one allows significantly larger NP contributions to $\Cc{7}$ and/or $\Cc{7'}$. However, the above exercise is enough to highlight that $\smash{\langle A_T^{(2)}\rangle_{\rm inc}}$ and $\smash{\langle A_T^{(\text{Im});CP}\rangle_{\rm inc}}$
are the most relevant observables to exploit mixing-induced effects to probe $b\to s\gamma$ Wilson coefficients, as will be considered in the following.
It also suggests that $\smash{\langle A_T^{(2);CP}\rangle_{\rm inc}}$ and $\smash{\langle A_T^{(\text{Im})}\rangle_{\rm inc}}$ (i.e.,~$\langle A_3\rangle_{\rm inc}$ and $\langle  S_9\rangle_{\rm inc}$) obtained from $d\Gamma-d\bar\Gamma$ can also be worth investigating as they yield a constraint on $\Cc{7}$ induced by neutral-meson mixing directly, as discussed in Appendix~\ref{app:asymmbis}.

\section{Numerical study}\label{sec:numerical}

\subsection{Inputs}

We consider the following inputs for our numerical studies:
\begin{itemize}
\item For the SM Wilson coefficients and their chirality-flipped counterparts, we take the values from Table 2 of Ref.~\cite{Altmannshofer:2008dz}, corresponding to a computation performed at the next-to-next-to-leading-logarithm accuracy at the scale $\mu_b=4.8$ GeV. In particular $\Cc{7}^{\rm SM}=-0.304$ and $\Cc{7'}^{\rm SM}= - 0.006$.
\item The NP scenarios considered are written as 
$\Cc{i}=\Cc{i}^{\rm SM}+\delta \Cc{i}$ at the same scale $\mu_b$.
\item For the CKM matrix elements, we take the most recent values of CKMfitter of spring 2021~\cite{CKMfitter} and symmetrise the 1$\sigma$ range for the quantities of interest, leading in particular to $\beta_s= 0.0185(4)$, $|V_{tb}| = 0.9991(2)$ and  $|V_{ts}| = 0.0406(5)$.
\item For the $B_s\to \phi$ form factors, we take the values of the light-cone sum rule analysis of Ref.~\cite{Bharucha:2015bzk}, taking into account the uncertainties 
reported in this article and, conservatively, assuming uncorrelated parameters. Similarly, whenever we need the $B\to K^\ast$ form factors, we consider the combination of light-cone sum rule and lattice results from Ref.~\cite{Bharucha:2015bzk} with uncorrelated parameters.
\item For the charm-loop contributions, we add to $\Cc{9}$ the perturbative function $Y(q^2)$ defined for instance in Ref.~\cite{Altmannshofer:2008dz}, with $m_c(m_c)= 1.27$ GeV and $m_b(m_b)=4.18$ GeV~\cite{Workman:2022ynf}. The soft-gluon contribution, not included in this function, can generate a contribution affecting $\Cc{7}$ and $\Cc{7'}$~\cite{Khodjamirian:2010vf}. Following the discussion in Ref.~\cite{Paul:2016urs}, we add a charm-loop contribution to the SM value with an arbitrary strong phase:
$\Cc{7^{(')}}^{\rm soft}= \rho^{(')} \exp(i \theta^{(')})$
    with $\theta,\theta'\in [0,\pi]$,
    $\rho\in [-1.5,1.5]\cdot 10^{-2}$ and  $\rho'\in [-0.4,0.4]\cdot 10^{-2}$ (with Gaussian uncertainties).
\end{itemize}

In the following, we will focus on the time-integrated measurements, as they are expected to be measured at LHCb~\cite{Desse:2020dta}. As indicated in Sec.~\ref{sec:timedepobs}, time-dependent measurements could provide a similar type of information on $\Cc{7}$ and $\Cc{7'}$ through the determination of the coefficients $s_3,s_9,h_3,h_9$, but they are likely to be difficult to attain in a near future. As a first illustration, we provide predictions of some $B_s\to \phi ee$ observables within the SM in Table~\ref{tab:SMpredictions1}. We illustrate the impact of mixing by showing the same observables neglecting mixing in Table~\ref{tab:SMpredictions2} and we provide similar observables for $B\to K^* ee$ observables  in Table~\ref{tab:SMpredictions3} without including any mixing effects, in agreement with the self-tagging $K^*$ decay considered in the LHCb measurement~\footnote{We have compared our predictions with the ones provided in {\tt flavio}~\cite{Straub:2018kue} for the observables implemented in their  framework. We have found reasonable agreement, with small differences that arise from the different values taken for the ``effective" SM Wilson coefficients $C_7$ and $C_9$.}. We only show the observables that are non-vanishing in the SM, using the bins considered for both reconstructed and effective $q^2$ values in Ref.~\cite{LHCb:2020dof}. We also consider bins with an upper value of 0.1 GeV$^2$, corresponding to a range closer to the photon pole, without making a distinction between reconstructed and effective $q^2$-values for the upper bound.

We note, in particular, that the mixing effects do modify the central values of the angular observables for $B_s\to\phi ee$. On the other hand, the exact position of the upper value of the bin has a very small impact on the value of the angular observables. This can be easily understood by the integration procedure used to compute the binned values: the angular observables are defined as ratios of angular coefficients $J_i$ (see Eqs.~(\ref{eq:defasymmetries}), (\ref{eq:defasymmbis}) and (\ref{eq:defasymmbis2}) out of which the normalization by the decay rate cancels), and the binned values of the angular observables are computed by taking the ratio of the integrated numerator and denominator in terms of these angular coefficients~\cite{
Descotes-Genon:2012isb,Descotes-Genon:2013vna
}. Both numerator and denominator are therefore dominated by the photon-pole contributions and exhibit little sensitive to the exact upper value of the bin.

\begin{table}[p]
\vspace{1cm}      
\centering
{\renewcommand{\arraystretch}{1.8}
\begin{tabular}{ |c|cccc| }
 \hline
    Decay mode  & \multicolumn{4}{c|}{$B_s\to \phi ee$}\\
\hline\hline
 $q^2$-bin~$[\text{GeV}^2]$ & $\langle\mathrm{Br} \rangle_{\rm inc} \times 10^{7}$ & $\langle F_L \rangle_{\rm inc} $ & $\langle A_T^{(2)}\rangle_{\rm inc}$ &  $\langle A_T^{(\mathrm{Im})}\rangle_{\rm inc}$   \\
 \hline
 $[0.0008,0.257]$ & $2.76 (26)$ & $ 0.106(19)$ & $0.111 (20)$ & $0.0369(2)$  \\
  $[0.0001,0.25]$ & $3.66(35)$ & $ 0.078(14) $ & $0.106(17)$ & $0.0369(2)$ \\
 $[0.0001,0.1]$ & $3.13(31)$ & $ 0.037(7) $ & $0.107(19)$ & $0.0369(2)$  \\
 \hline
\end{tabular} }
\caption{\sl\small SM predictions for $B_s\to \phi ee$ observables in the low-$q^2$ region. Note that the branching fraction corresponds to the sum of $\mathrm{Br}(B_s\to\phi ee)$ and $\mathrm{Br}(\overline{B}_s\to\phi ee)$.
}
\label{tab:SMpredictions1}
\end{table}
\begin{table}
\vspace{1cm}      
\centering
{ \renewcommand{\arraystretch}{1.8}
\begin{tabular}{ |c|cccc| }
 \hline
    Decay mode  & \multicolumn{4}{c|}{$B_s\to \phi ee$ (mixing not included)}\\
\hline\hline
 $q^2$-bin~$[\text{GeV}^2]$ & $\langle\mathrm{Br} \rangle_{\rm inc} \times 10^{7}$ & $\langle F_L \rangle_{\rm inc} $ & $\langle A_T^{(2)}\rangle_{\rm inc}$ &  $\langle A_T^{(\mathrm{Im})}\rangle_{\rm inc}$  \\
 \hline
 $[0.0008,0.257]$ & $2.79 (25)$ & $ 0.113(19)$ & $0.042 (20)$ & $0.00(2)$  \\
  $[0.0001,0.25]$ & $3.69(34)$ & $ 0.084(14) $ & $0.040(19)$ & $0.00(2)$  \\
 $[0.0001,0.1]$ & $3.15(30)$ & $ 0.039(7) $ & $0.040(19)$ & $0.00(2)$  \\
$[0.0008,0.1]$ & $2.23(21)$ & $ 0.055(10) $ & $0.040(19)$ & $0.00(2)$ \\
 \hline
\end{tabular} }
\caption{\sl\small SM predictions for $B_s\to \phi ee$ observables without including the effect of mixing in the low-$q^2$ region. Note that the branching fraction corresponds to the sum of $\mathrm{Br}(B_s\to\phi ee)$ and $\mathrm{Br}(\overline{B}_s\to\phi ee)$.}
\label{tab:SMpredictions2}
\end{table}
\begin{table}
\vspace{1cm}      
\centering
{ \renewcommand{\arraystretch}{1.8}
\begin{tabular}{ |c|cccc| }
 \hline
    Decay mode  &\multicolumn{4}{c|}{$B\to K^* ee$}\\
\hline\hline
 $q^2$-bin~$[\text{GeV}^2]$ & $\langle\mathrm{Br} \rangle_{\rm inc} \times 10^{7}$ & $\langle F_L \rangle_{\rm inc} $ & $\langle A_T^{(2)}\rangle_{\rm inc}$ &  $\langle A_T^{(\mathrm{Im})}\rangle_{\rm inc}$  \\
 \hline
 $[0.0008,0.257]$ & $1.40(26) $ & $ 0.087(21) $ & $0.04(2)$ & $0.00(2)$\\
  $[0.0001,0.25]$ & $1.87(35)$ & $ 0.063(16) $ & $0.04(2)$ & $0.00(2)$\\
 $[0.0001,0.1]$ & $1.61(31)$ & $ 0.030(8) $ & $0.04(2)$ & $0.00(2)$\\
$[0.0008,0.1]$ &  $1.14(22)$ & $0.042(10) $ & $0.04(2)$ & $0.00(2)$\\
 \hline
\end{tabular} }
\caption{\sl\small SM predictions for $B_d\to K^{*\,0}ee$ observables in the low-$q^2$ region.  Mixing effects are not included in this case since they are not relevant for $B \to K^{*0}(\rightarrow K^+\pi^-)ee$ decays.} 
\label{tab:SMpredictions3}
\end{table}

\begin{table}[p]
\vspace{1cm}      
\centering
\renewcommand{\arraystretch}{1.35}
\begin{tabular}{ |p{1.8cm}|p{1.2cm}|p{1.2cm}|p{1.2cm}| }
 \hline
 \multicolumn{4}{|c|}{{Relative Difference}} \\
 \hline\hline
 $q^2~[\text{GeV}^2]$  & $0.05$&  $0.15$ &  $0.3$  \\
 \hline
 $d\text{Br}/dq^2$   & $8 \% $ & $20 \% $ &  $ 34 \%$  \\
 \hline
 $S_{1s}$ & $12 \% $ & $30 \% $ & $ 49 \%$   \\
 $S_{2s}$ & $12 \% $ & $30 \% $ & $ 49 \%$   \\
 $S_{3}$    & $9 \% $ & $ 23 \% $ &  $38 \%$  \\
 $S_{9}$ & $12 \% $ & $30 \% $ & $ 49 \%$  \\
 $A_T^{(2)}$ & $4 \% $ & $11 \% $ & $ 21 \%$  \\
 $A_{1s}$ & $6 \% $ & $15 \% $ &  $ 28 \%$  \\
 $A_{2s}$ & $6 \% $ & $15 \% $ &  $ 28 \%$  \\
 $A_{3}$ & $5 \% $ & $15 \% $ &  $ 27 \%$  \\
 $A_{9}$ & $6 \% $ & $16 \% $ &  $ 29 \%$  \\
 $A_T^{(\text{Im});CP}$  & $0.4 \% $ & $2 \% $ &  $ 4 \%$  \\
 \hline
\end{tabular} 
\caption{\sl\small Relative difference between the complete expression for the observables and the photon-pole approximation for three different values of $q^2 \in [4m_e^2, 0.3~\mathrm{GeV}^2]$. For $d\text{Br}/dq^2$, $S_{1s}$, $S_{2s}$, $S_{3}$, $S_{9}$ and $A_T^{(2)}$ we consider $\delta\Cc{7} = \delta\Cc{7'} = 0$, while for $A_{1s}$, $A_{2s}$, $A_{3}$, $A_{9}$ and $A_T^{(\text{Im});CP}$ we consider $(\delta\Cc{7},\delta\Cc{7'}) = (0.2i,0.3i)$. 
}
\label{tab:photonpoleapprox}
\end{table}
\begin{table}[]
\vspace{1cm}      
\centering
\renewcommand{\arraystretch}{1.5}
\begin{tabular}{ |p{2cm}|p{2.05cm}|p{2.05cm}|p{1.87cm}| }
 \hline
 \multicolumn{4}{|c|}{{Relative Difference}} \\
 \hline\hline
 $q^2$~[$\text{GeV}^2$]  & $[0.0008, 0.05]$&  $[0.0008, 0.15]$ &  $[0.0008, 0.3]$  \\
 \hline
 $\langle \text{Br} \rangle_{\rm inc}$ & $2 \% $ & $5 \% $ & $ 8 \%$   \\
 \hline
 $\langle S_{1s} \rangle_{\rm inc}$ & $3 \% $ & $7 \% $ & $ 13 \%$   \\
 $\langle S_{2s} \rangle_{\rm inc}$ & $3 \% $ & $7 \% $ & $ 13 \%$   \\
 $\langle S_{3} \rangle_{\rm inc}$    & $2 \% $ & $ 5 \% $ &  $9 \%$  \\
 $\langle S_{9} \rangle_{\rm inc}$ & $3 \% $ & $7 \% $ & $ 13 \%$  \\
 $\langle A_T^{(2)} \rangle_{\rm inc}$ & $1 \% $ & $2 \% $ & $ 3 \%$  \\
 $\langle A_{1s} \rangle_{\rm inc}$ & $1 \% $ & $3 \% $ &  $ 5 \%$  \\
 $\langle A_{2s} \rangle_{\rm inc}$ & $1 \% $ & $3 \% $ &  $ 5 \%$  \\
 $\langle A_{3} \rangle_{\rm inc}$ & $1 \% $ & $3 \% $ &  $ 5 \%$  \\
 $\langle A_{9} \rangle_{\rm inc}$ & $1 \% $ & $3 \% $ &  $ 6 \%$  \\
 $\langle A_T^{(\text{Im});CP} \rangle_{\rm inc}$  & $0.1 \% $ & $0.3 \% $ &  $0.5 \%$ \\
 \hline
\end{tabular} 
\caption{\sl\small Relative difference between the mean value for the observables and the photon-pole approximation for three different bins of $q^2 \in [0.0008, 0.05]~\mathrm{GeV}^2$, $[0.0008, 0.15]~\mathrm{GeV}^2$ and $[0.0008, 0.3]~\mathrm{GeV}^2$ . For $\text{Br}$, $S_{1s}$, $S_{2s}$, $S_{3}$, $S_{9}$ and $A_T^{(2)}$ we consider $\delta\Cc{7} = \delta\Cc{7'} = 0$, while for $A_{1s}$, $A_{2s}$, $A_{3}$, $A_{9}$ and $A_T^{(\text{Im});CP}$ we consider $(\delta\Cc{7},\delta\Cc{7'}) = (0.2i,0.3i)$. 
}
\label{tab:photonpoleapprox-bix}
\end{table}

\subsection{SM predictions and validity of the photon-pole approximation}\label{sec:validityphotonpole}

In Sec.~\ref{sec:Bsnomixing}, we considered the photon-pole approximation and we argued that this approximation should be valid for $q^2$ in the lower part of the experimental range currently probed. We are now in a position to check this by considering the value of the observables of interest in the $q^2$ range between $4m_e^2$ and $0.3$ GeV$^2$, and comparing the value obtained with the contributions from all amplitudes and the value in the photon-pole approximation discussed in Sec.~\ref{sec:validityphotonpole}.

\begin{figure}[!p]
\begin{center}
\hspace{-3.3em}\includegraphics[width=.52\linewidth]{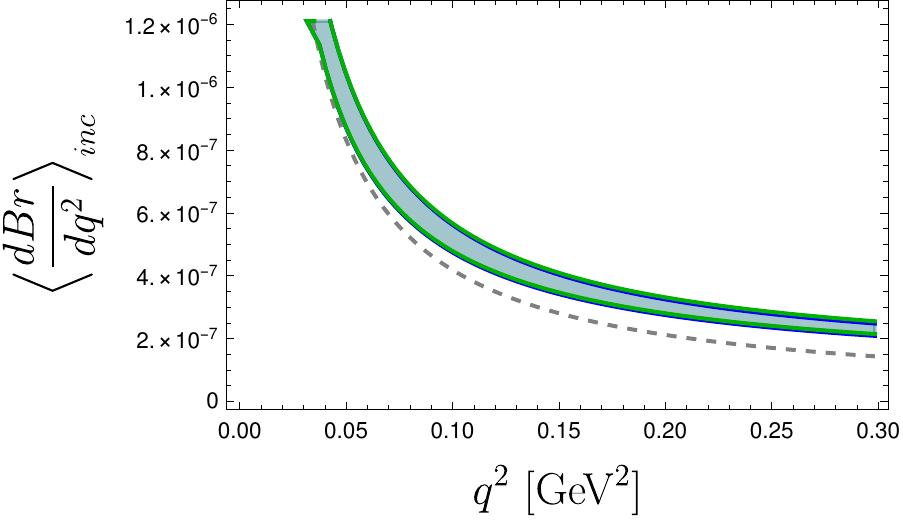}
\includegraphics[width=.46\linewidth]{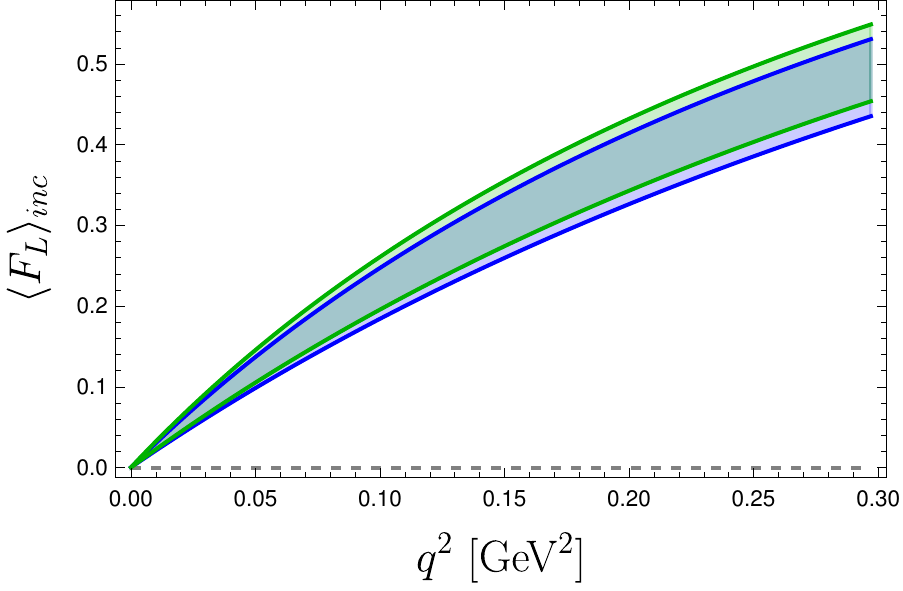} 
\\[0.5em]
\includegraphics[width=.46\linewidth]{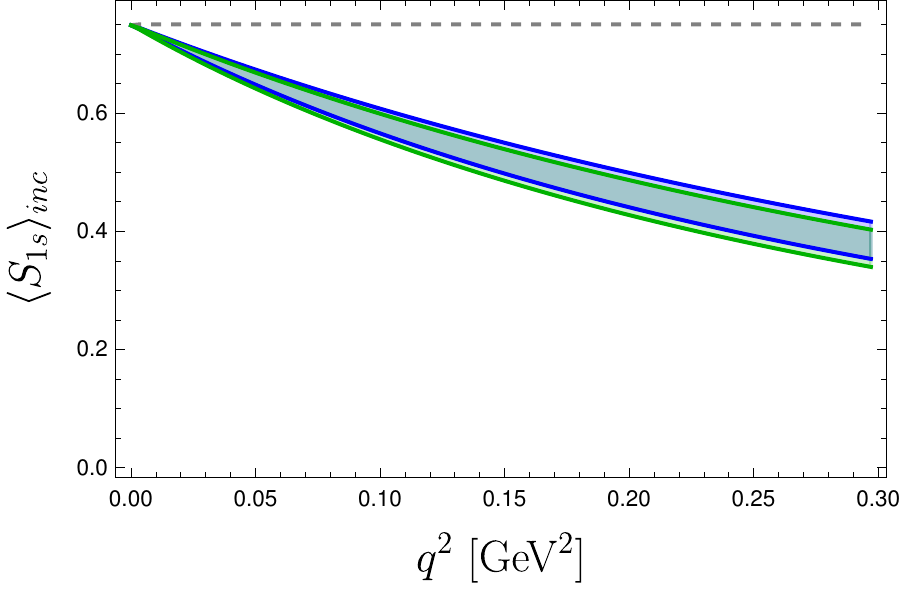}
\includegraphics[width=.46\linewidth]{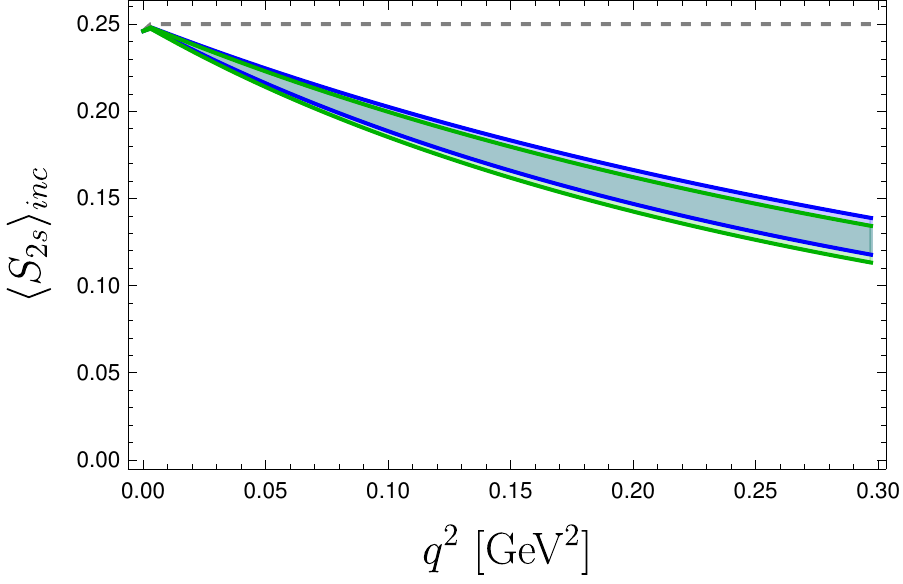} 
\\[0.5em]
\includegraphics[width=.46\linewidth]{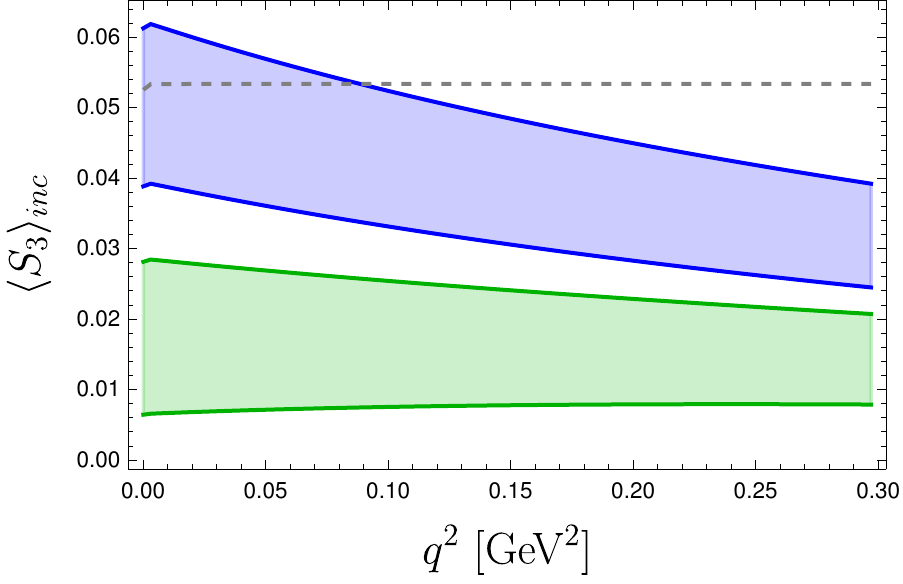}
\includegraphics[width=.46\linewidth]{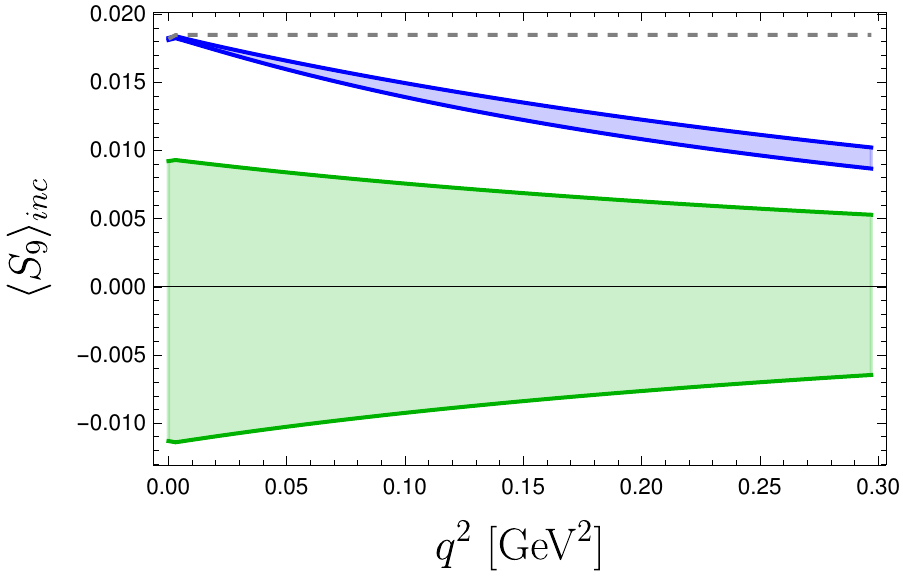} 
\\[0.5em]
\hspace{3cm}\includegraphics[width=.65\linewidth]{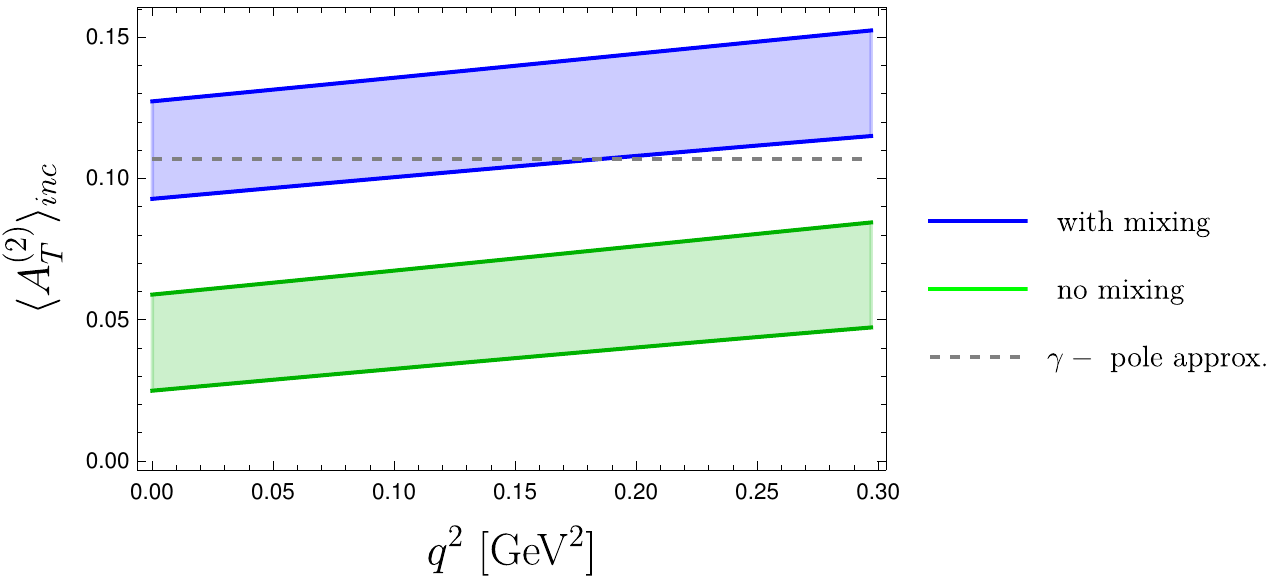}

\end{center}
\caption{\sl\small SM predictions for $d{\rm Br}/dq^2$, and the angular observables $F_L$, $S_{1s}$, $S_{2s}$, $S_3$, $S_9$, and $A_T^{(2)}$, as a function of $q^2$. Predictions are shown for the complete expression (blue) and the photon-pole approximation (dashed gray). The full result in the absence of mixing is also shown for illustration (green).
\label{fig:photonpole1}
}
\end{figure}

\begin{figure}[!p]
\begin{center}
\includegraphics[width=.46\linewidth]{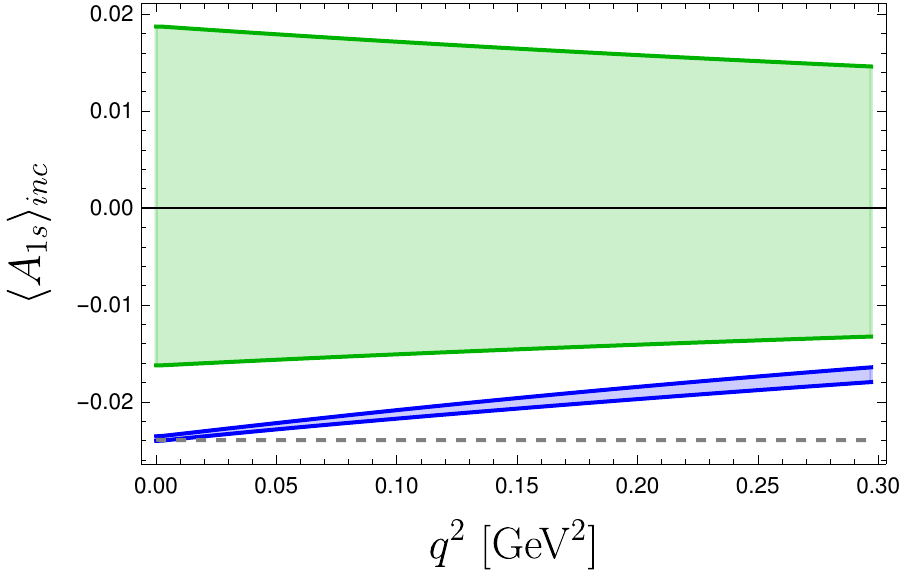} \hspace{1cm}
\includegraphics[width=.46\linewidth]{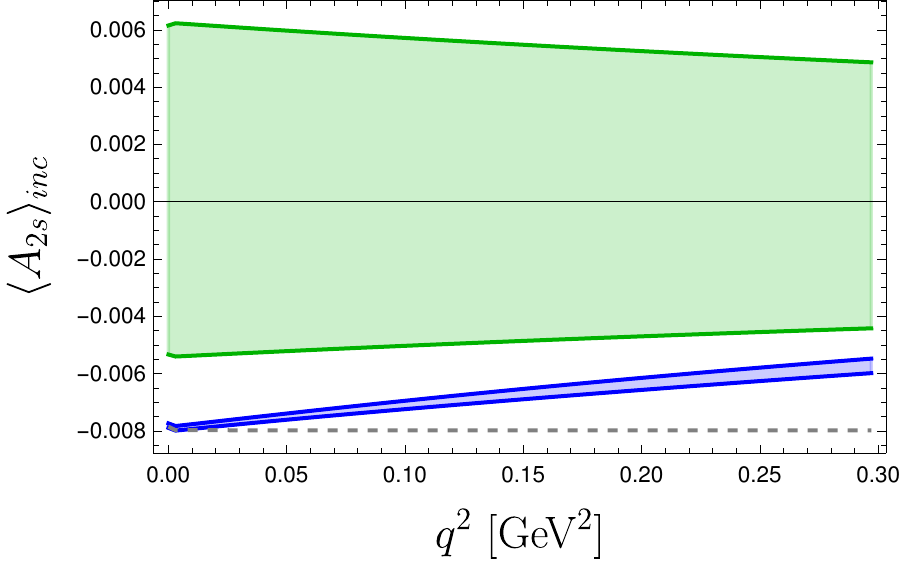}
\\
\vspace{1cm}
\includegraphics[width=.46\linewidth]{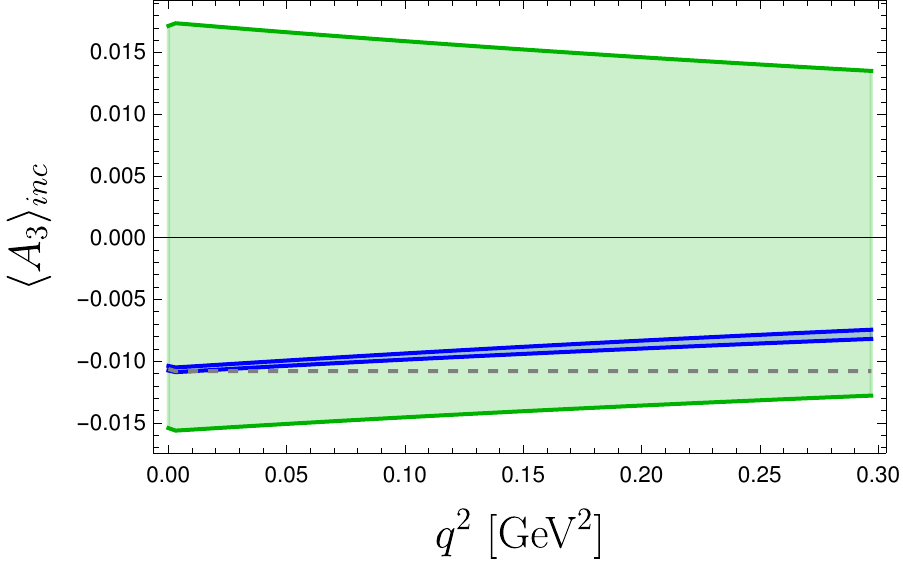} \hspace{1cm}
\includegraphics[width=.46\linewidth]{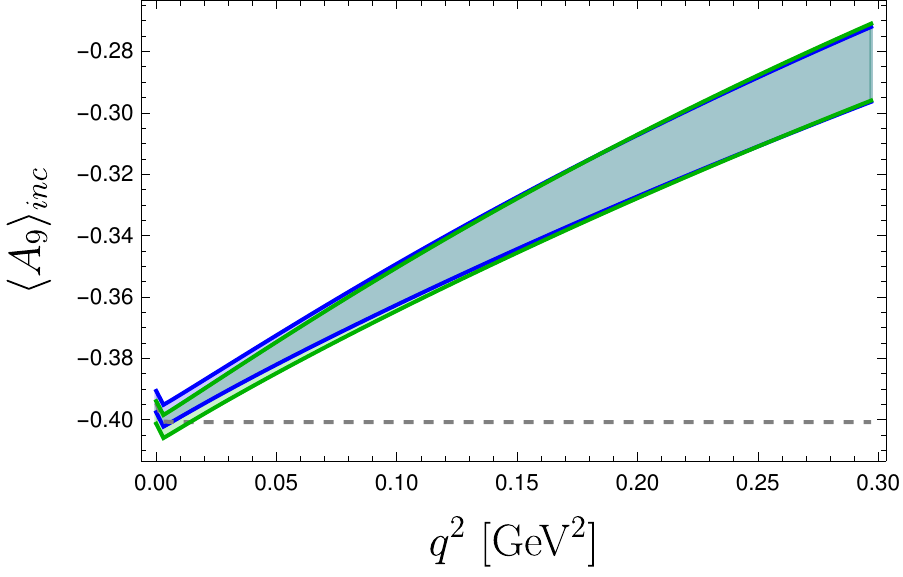}
\\
\vspace{1cm}
\hspace{3cm}\includegraphics[width=.65\linewidth]{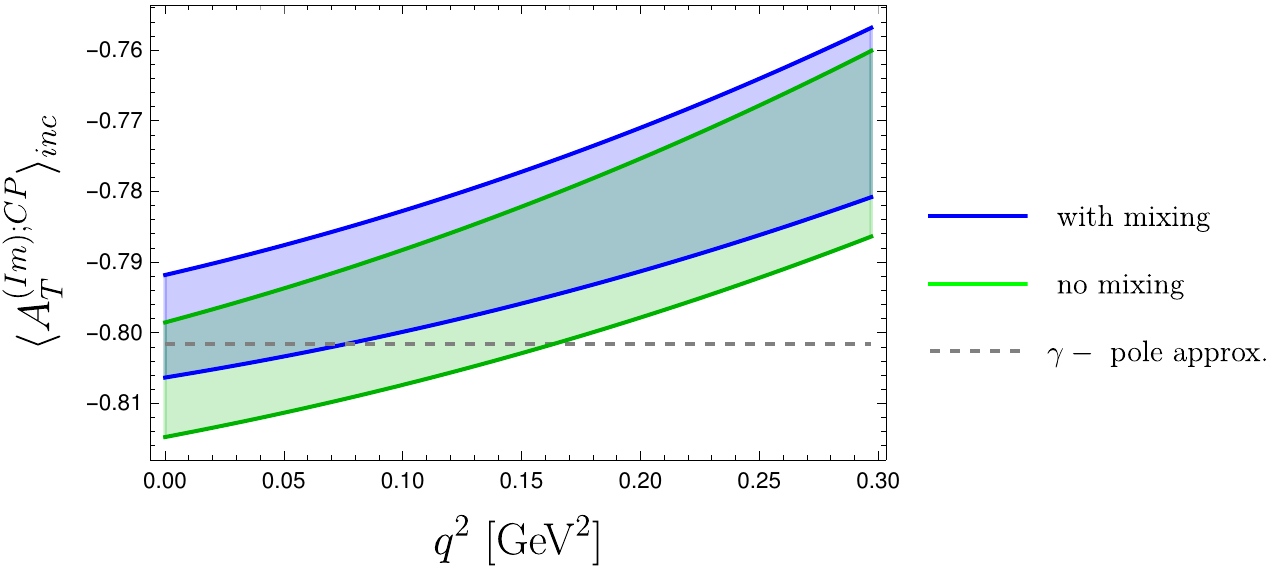}
\end{center}
\caption{\sl\small Predictions for $A_{1s}$, $A_{2s}$, $A_3$, $A_9$ and $A_T^{(\text{Im});CP}$ as a function of  $q^2$ in the NP scenario defined by $(\delta\Cc{7}, \delta\Cc{7'}) = (0.2i,0.3i)$. Predictions are shown for the complete expression (blue) and the photon-pole approximation (dashed gray). The full result in the absence of mixing is also shown for illustration (green).
\label{fig:photonpole2}
}
\end{figure}

In Fig.~\ref{fig:photonpole1}, we predict the differential branching fractions, as well as the angular observables $S_{1s}$, $S_{2s}$, $S_3$, $S_9$ and $A_T^{(2)}$ as a function of $q^2$, using the SM values for the Wilson coefficients. Similar predictions are given in Fig.~\ref{fig:photonpole2}  for $A_{1s}$, $A_{2s}$, $A_3$, $A_9$ and $\smash{A_T^{(\text{Im});CP}}$ where we take $(\delta\Cc{7},\delta\Cc{7'}) = (0.2i,0.3i)$, since these CP-asymmetries vanish in the absence of NP weak phases. In these plots, we display the complete expressions in the presence (blue) and absence of mixing (green), in addition to the photon-pole approximation depicted by the dashed gray line. 

The relative difference between the complete and photon-pole expressions at the differential leval is collected in Table~\ref{tab:photonpoleapprox}, for the observables where the latter does not vanish. One can see that the agreement is reasonably good between the exact expression and the photon-pole approximation below 0.05 GeV$^2$ and that it deteriorates for larger $q^2$ values, reaching up to $\approx 50$\% for some of the observables at $q^2=0.3$ GeV$^2$.
 The photon-pole approximation can be a useful starting point to understand the constraints on $\Cc{7}$ and $\Cc{7'}$ extracted from low-$q^2$ observables. However, it is also clear that the approximation cannot be held as accurate beyond 0.2 GeV$^2$ at the differential level, as it can also be seen in Fig.~\ref{fig:photonpole1}.  For the CP-averaged quantities, we see that the observables $S_3, S_9$ and $A_T^{(2)}$ are quite affected by mixing effects, shifting the central values significantly. In the case of CP asymmetries, for the NP scenario chosen for illustration, we see that some of the observables are quite affected as well, since they are actually dominated by mixing effects, which explains the dramatic change in the uncertainties in Fig.~\ref{fig:photonpole1} once neutral-meson mixing is considered. Note, also, that the asymmetry
$A_T^{(\text{Im});CP}$ is only mildly affected in comparison to the others.

Lastly, we stress that the agreement between the full expressions and the photon-pole dominance approximation is considerably improved for binned observables at low-$q^2$, as shown in Table~\ref{tab:photonpoleapprox-bix}. This is the case since the photon pole dominates the integrals involved in the binned observables, which implies that this approximation has a larger range of applicability for binned observables.

\subsection{Constraints on $\Cc{7},\Cc{7'}$ scenarios}

The transverse asymmetries considered by LHCb for $B\to K^*ee$ with 9 fb$^{-1}$~\cite{LHCb:2020dof} led to statistical uncertainties of about $10\%$, dominating over the systematic ones, and considered in a $q^2$ range between 0.0008 and 0.257 GeV$^2$. We expect the $B_s\to \phi ee$ analysis to be of similar complexity, even though some features might help the analysis, such as the narrow width of the $\phi$ meson compared to the $K^*$ and the heavier mass of the $B_s$ meson. On the other hand, LHCb produces approximately 4 times more $B_d$ mesons than $B_s$ since $f_s/f_d \sim 1/4$~\cite{Desse:2020dta}. All in all, one could expect to reach uncertainties of similar sizes for the observables measured in both modes.

\begin{figure}[!t]
\begin{center}
\includegraphics[width=0.48\textwidth]{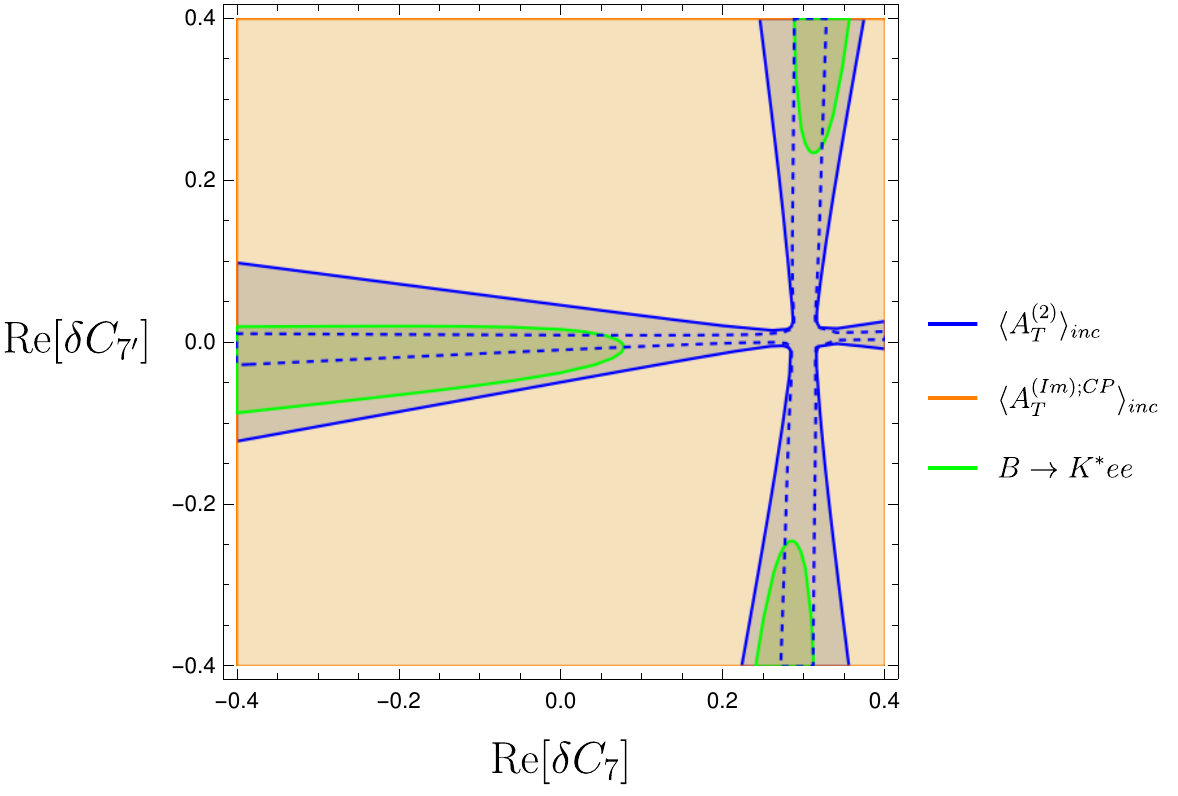} 
\includegraphics[width=0.48\textwidth]{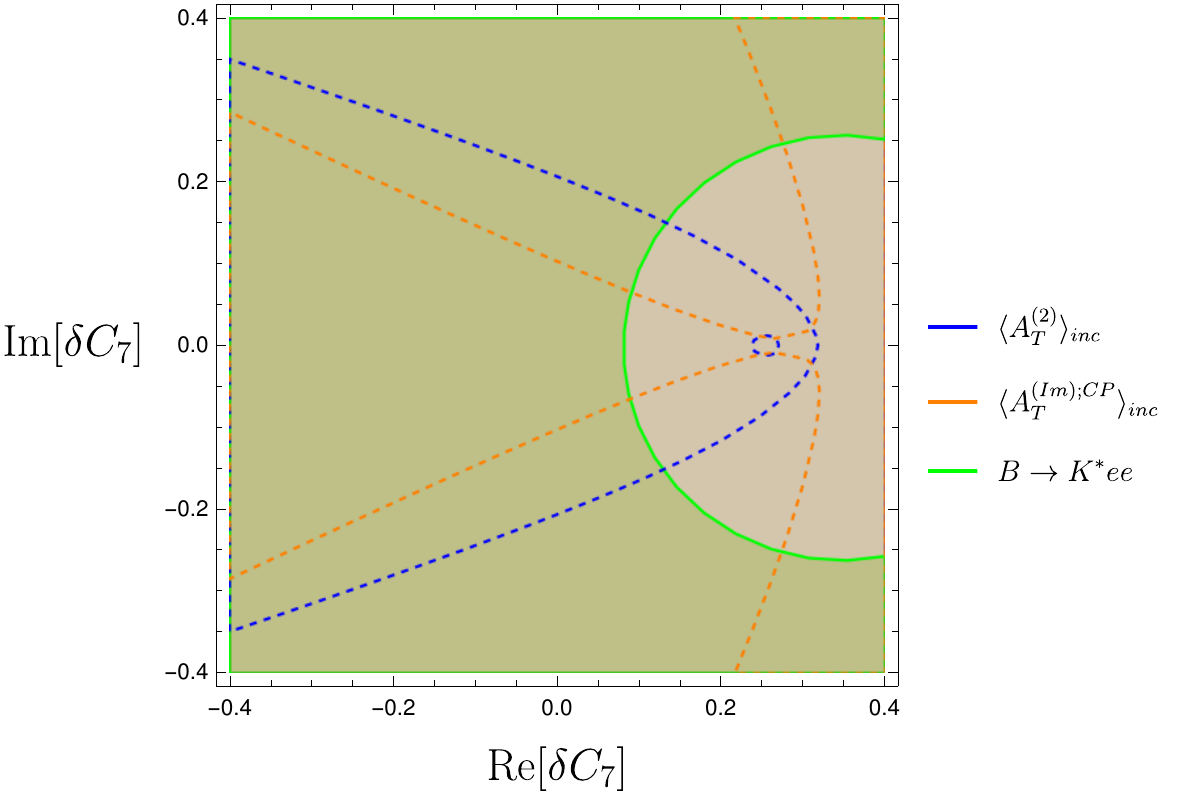} 
\\
\vspace{1cm}
\includegraphics[width=0.48\textwidth]{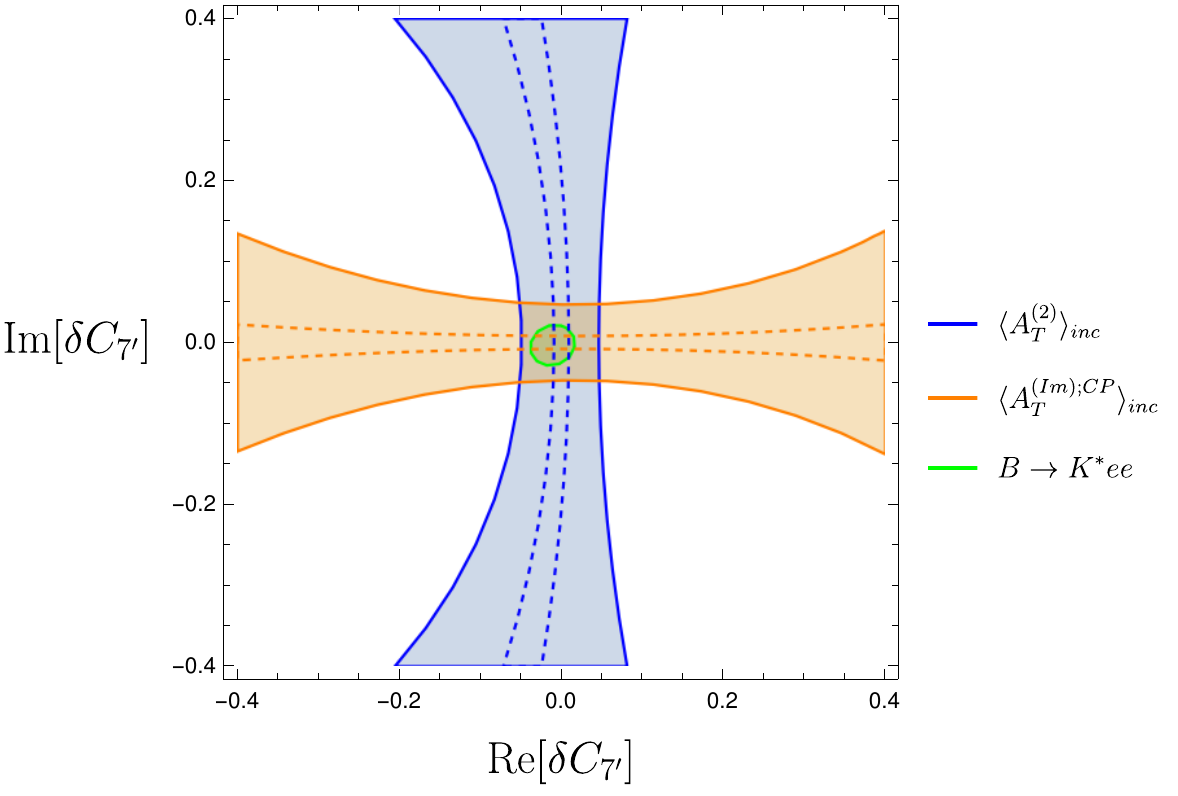} 
\includegraphics[width=0.48\textwidth]{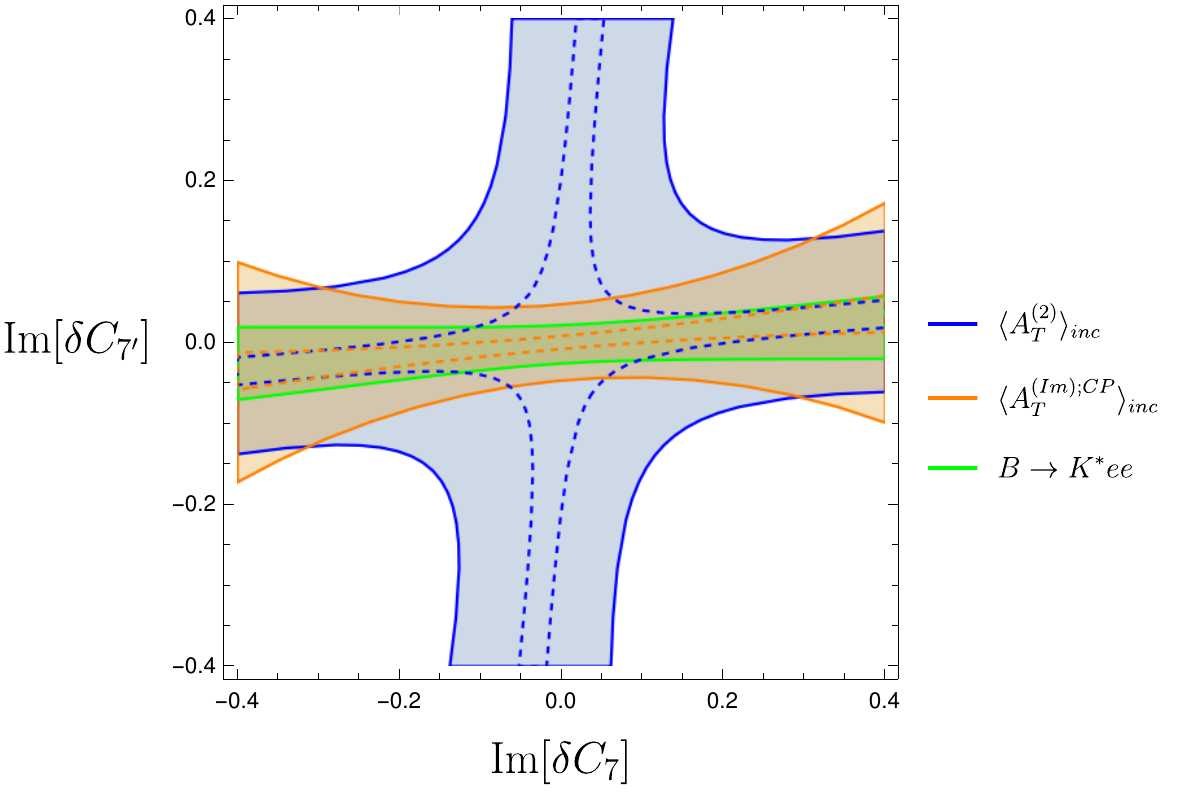}
\end{center}
\caption{\sl\small Projected 1$\sigma$ contours on NP contributions to $\Cc{7}$ and $\Cc{7'}$ from $B_s\to\phi ee$ observables at low $q^2$: 
 $\langle A_T^{(2)}\rangle_{\rm inc} $ and $\langle A_T^{(\text{Im});CP}\rangle_{\rm inc}$.
Four NP scenarios are shown: (i) ${\rm Re}(\Cc{7})$ vs.~${\rm Re}(\Cc{7'})$, 
(ii) ${\rm Re}(\Cc{7})$ vs.~${\rm Im}(\Cc{7})$, 
(iii) ${\rm Re}(\Cc{7'})$ vs.~${\rm Im}(\Cc{7'})$ and (iv)
${\rm Im}(\Cc{7})$ vs.~${\rm Im}(\Cc{7'})$. 
The SM point corresponds to the origin (0,0). Meson mixing is taken into account for $B_s\to \phi ee$ observables. The dashed lines correspond to the future projections (300 $\text{fb}^{-1}$) for $A_T^{(2)}$ (blue) and $A_T^{(\text{Im});CP}$ (orange). 
\label{fig:NPcontours}
}
\end{figure}

\begin{figure}[!t]
\begin{center}
{\includegraphics[width=0.48\textwidth]{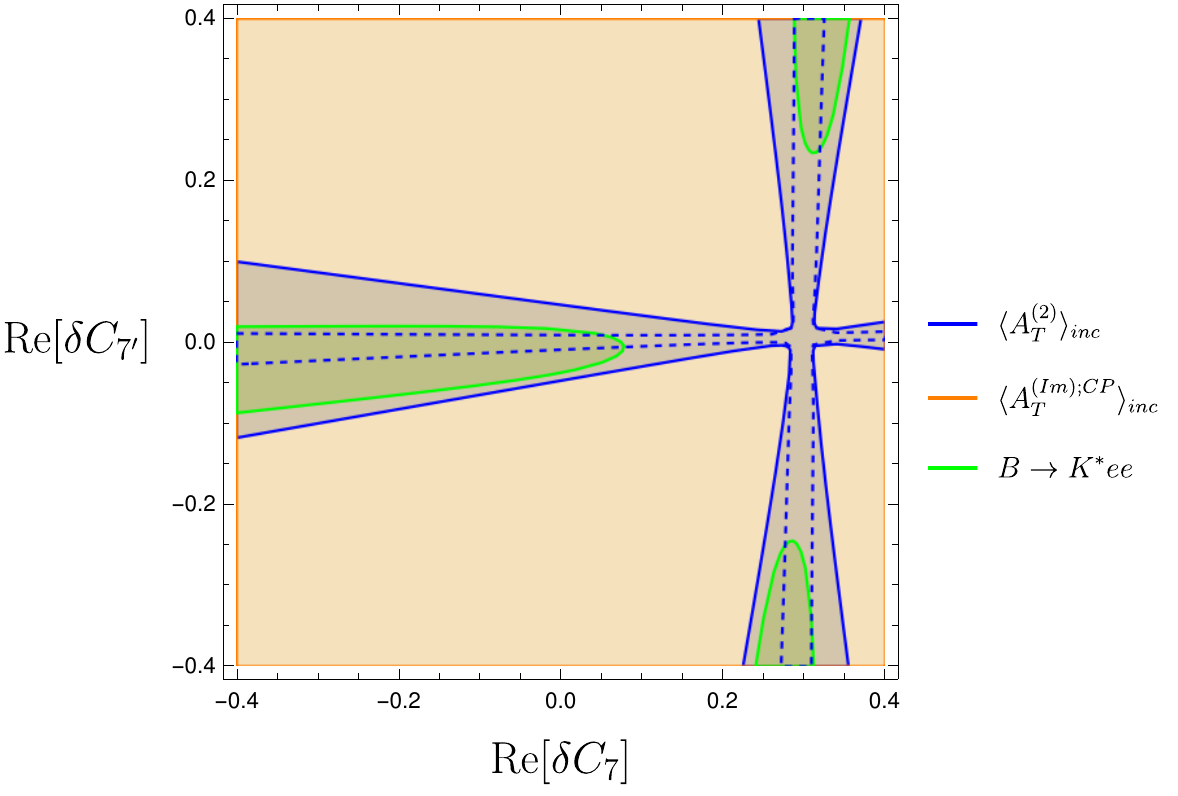} 
\put(-179,139) {\footnotesize\color{gray} no mixing}}
{\includegraphics[width=0.48\textwidth]{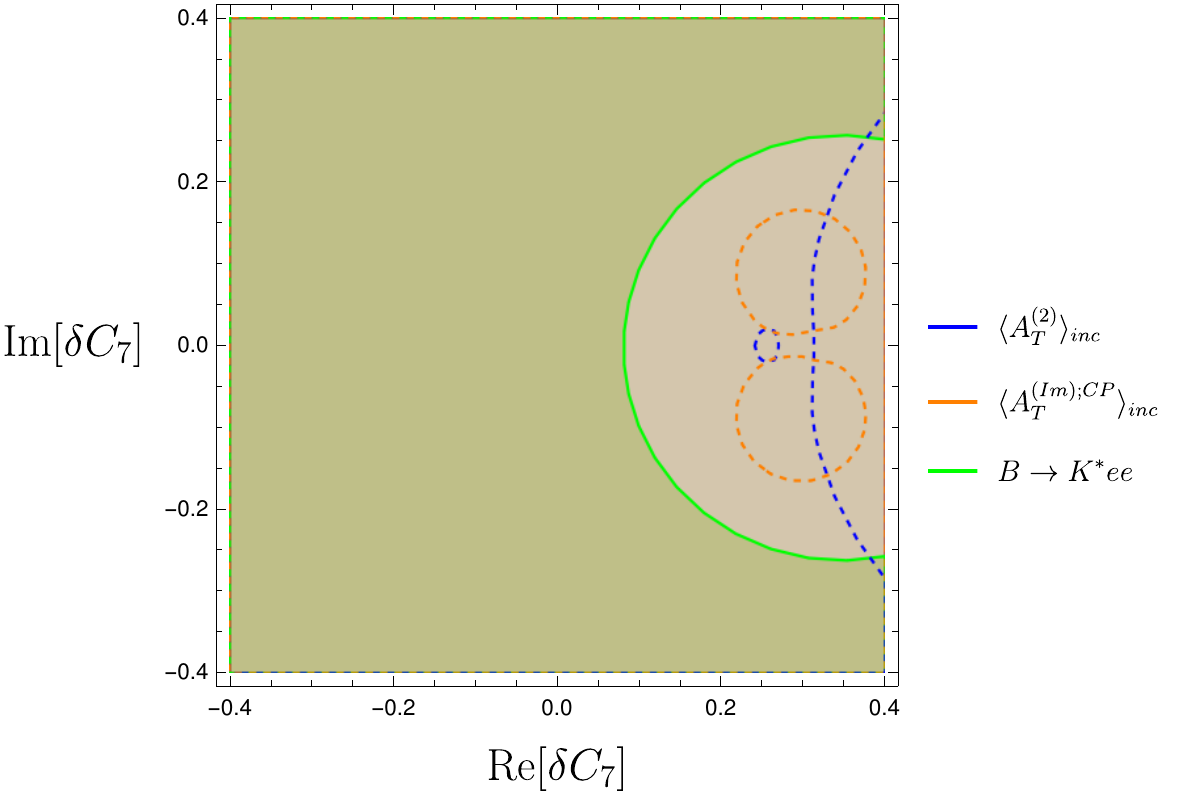} 
\put(-179,139) {\footnotesize\color{gray} no mixing}}
\\
\vspace{1cm}
{\includegraphics[width=0.48\textwidth]{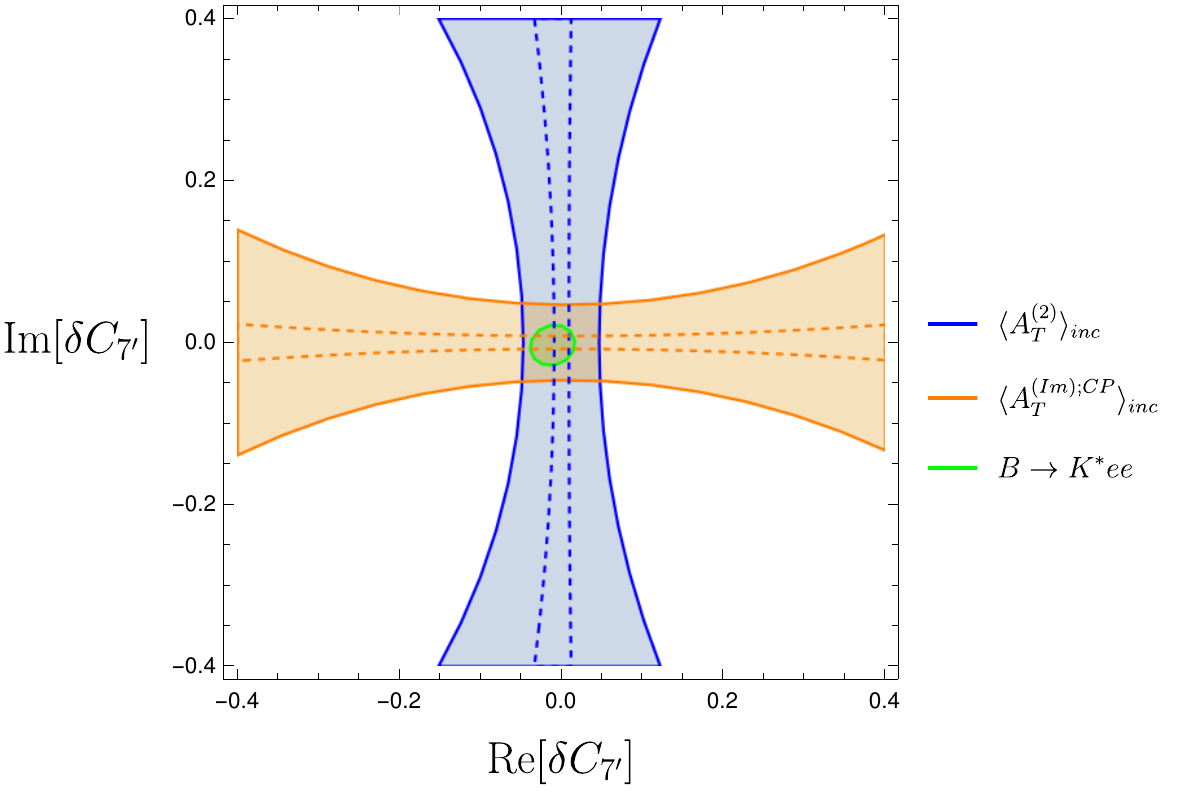} 
\put(-179,139) {\footnotesize\color{gray} no mixing}}
{\includegraphics[width=0.48\textwidth]{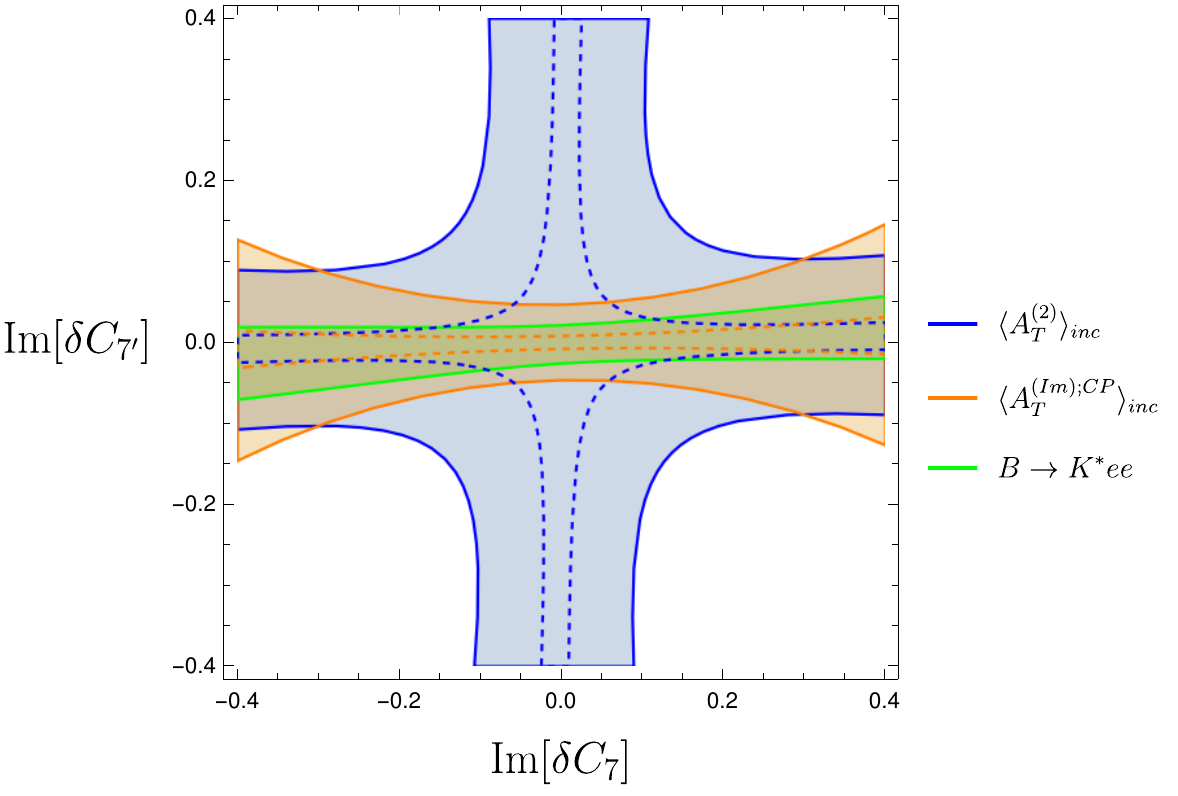}
\put(-179,139) {\footnotesize\color{gray} no mixing}}
\end{center}
\caption{\sl\small Projected 1$\sigma$ contours on NP contributions to $\Cc{7}$ and $\Cc{7'}$ from $B_s\to\phi ee$ observables at low $q^2$: 
 $\langle A_T^{(2)}\rangle_{\rm inc} $ and $\langle A_T^{(\text{Im});CP}\rangle_{\rm inc}$.
Four NP scenarios are shown: (i) ${\rm Re}(\Cc{7})$ vs.~${\rm Re}(\Cc{7'})$, 
(ii) ${\rm Re}(\Cc{7})$ vs.~${\rm Im}(\Cc{7})$, 
(iii) ${\rm Re}(\Cc{7'})$ vs.~${\rm Im}(\Cc{7'})$ and (iv)
${\rm Im}(\Cc{7})$ vs.~${\rm Im}(\Cc{7'})$. 
The SM point corresponds to the origin (0,0).  
For illustration, meson mixing is not taken into account for $B_s\to \phi ee$ observables. The dashed lines correspond to the future projections (300 $\text{fb}^{-1}$) for $A_T^{(2)}$ (blue) and $A_T^{(\text{Im});CP}$ (orange). 
\label{fig:NPcontours-nomixing}
}
\end{figure}

More specifically, we will focus on the angular asymmetries that can be derived from $d\Gamma+d\bar\Gamma$ (see Appendix~\ref{app:asymmbis} for the other ones).
We consider  four two-dimensional scenarios, in the planes ${\rm Re}(\Cc{7})$ vs.~${\rm Re}(\Cc{7'})$, 
${\rm Re}(\Cc{7})$ vs.~${\rm Im}(\Cc{7})$, ${\rm Re}(\Cc{7'})$ vs.~${\rm Im}(\Cc{7'})$, and ${\rm Im}(\Cc{7})$ vs.~${\rm Im}(\Cc{7'})$, respectively.
With this choice of scenarios, we take two transverse asymmetries 
$\langle A_T^{(2)}\rangle_{\rm inc} $ and $\langle A_T^{(\text{Im});CP}\rangle_{\rm inc}$ in the $q^2$ bin between 0.0008 and 0.257 GeV$^{2}$, and we assume that these observables have been measured with a central value equal to the SM expectation (including mixing effects) and a projected uncertainty of 0.2~\footnote{Note that if we consider more extensive NP scenarios with additional weak phases in $\Cc{9^{(')}}$ and $\Cc{10^{(')}}$, other observables could become very relevant, in particular $\smash{A_T^{(\text{Re});CP}}$, but also the other angular observables defined in Eq.~(\ref{eq:dist}).}. Furthermore, we consider a naive projection to 300 $\mathrm{fb}^{-1}$ corresponding to the LHCb second upgrade, where we rescale the uncertainties assuming that they are statistically dominated, see e.g.~Ref.~\cite{Desse:2020dta}.

The resulting constraints are shown in Fig.~\ref{fig:NPcontours},
which can be understood in terms of the simplified expressions given in Eqs.~(\ref{eq:AT2nearSM})-(\ref{eq:ATimCPnearSM}), focusing first on the terms independent from meson mixing.
 Due to the presence of the real SM contribution to $\Cc{7}$, the observable $\smash{\langle A_T^{(\text{Im});CP}\rangle_{\rm inc}}$  mainly provides constraints for NP scenarios with imaginary parts for $\Cc{7}$ and $\Cc{7'}$.
Moreover, one expects indeed 
a cross (hyperbolic) shape from $\smash{\langle A_T^{(2)}\rangle_{\rm inc}}$ 
for the scenarios involving only real parts or only imaginary parts of the Wilson coefficients, see Eq.~\eqref{eq:AT2nearSM}. For the scenario with NP only in $\Cc{7'}$, the presence of a large real part in $\Cc{7}$
constraints the real  part of  $\Cc{7'}$, whereas
 the absence of a large imaginary part in $\Cc{7}$ leaves the imaginary part of  $\Cc{7'}$ unconstrained,
which leads to the vertical band observed for $\langle A_T^{(2)}\rangle_{\rm inc} $, see Eq.~\eqref{eq:ATimnearSM}.
Similar arguments can be derived
to explain the bands observed in the case of $\smash{\langle A_T^{(\text{Im});CP}\rangle_{\rm inc}}$ for NP scenarios affecting either ${\rm Re}(\Cc{7'})$ and ${\rm Im}(\Cc{7'})$, or
${\rm Im}(\Cc{7})$ and ${\rm Im}(\Cc{7'})$, and to explain the absence of sensitivity for NP in ${\rm Re}(\Cc{7})$ and ${\rm Im}(\Cc{7})$.

For comparison, we also indicate for each scenario the region derived from current measurements on $B\to K^*ee$~\cite{LHCb:2015ycz,LHCb:2020dof} concerning $F_L$, $\smash{A_T^{(2)}}$, $A_T^{\rm Re}$ and $A_T^{\rm Im}$. We recall that the experimental notation $A_T^{\rm Im}$ for $B\to K^*ee$~\cite{LHCb:2015ycz,LHCb:2020dof} corresponds actually to a CP-asymmetry and would be denoted $\smash{A_T^{({\rm Im});CP}}$ in our convention. The last two observables are driving the constraints shown in Fig.~\ref{fig:NPcontours} (and drawn again in Ref.~\ref{fig:NPcontours-nomixing}). Clearly, $B\to K^*ee$ observables are measured using a self-tagging mode, which means that no mixing effects are included in their theoretical computations. The measured central values are close to the SM expectations, and the uncertainties are slightly smaller than the ones chosen for $B_s\to \phi ee$, which explains the good compatibility between the contours obtained from the two modes for most of the scenarios. The region excluded by $B\to K^*ee$ measurements in the case of NP in $\mathrm{Re}(\Cc{7})$ vs.~$\mathrm{Im}(\Cc{7})$ (upper right panel) comes from $F_L(B\to K^*ee)$.

Lastly, since we explored the impact of $B_s$ meson mixing for the observables considered above, it is instructive to display the same plots in the absence of meson mixing (i.e.~setting $x$ and $y$ to zero in the above expressions).
The resulting constraints are shown in Fig.~\ref{fig:NPcontours-nomixing}. In particular, one can see that the shapes are modified in the top-right and lower-right panels, indicating that mixing effects have a non-negligible impact on the constraints derived from these asymmetries. Furthermore, we reiterate that the central values of specific SM observables can also be significantly shifted, as shown e.g.~for $A_T^{(2)}$ in Fig.~\ref{fig:photonpole1}.

\subsection{Sensitivity to New Physics weak phases}

Finally we explore the sensitivity of the various observables to complex NP couplings, by considering the following NP scenarios: 
\begin{itemize}
\item SM case: $(\delta\Cc{7},\delta\Cc{7'})= (0,0)$\,;
\item Real $\delta\Cc{7}$\,, i.e.~real shift in $\Cc{7}$: $(\delta\Cc{7},\delta\Cc{7'})=\Cc{7}^{\rm SM}(0.2,0)$\,;
\item Real $\delta\Cc{7'}$\,, i.e.~real shift in $\Cc{7'}$: $(\delta\Cc{7},\delta\Cc{7'})=\Cc{7}^{\rm SM}(0,0.2)$\,;
\item Complex $\delta\Cc{7}$\,, i.e.~complex shift in  $\Cc{7}$: $(\delta\Cc{7},\delta\Cc{7'})=\Cc{7}^{\rm SM}(0.2  e^{i\theta},0)$\,;
\item  Complex $\delta\Cc{7'}$\,, i.e.~complex shift in  $\Cc{7'}$: $(\delta\Cc{7},\delta\Cc{7'})=\Cc{7}^{\rm SM}(0,0.3e^{i\theta})$\,;
\item  Correlated case, i.e.~correlated complex shift in $\Cc{7}$  and $\Cc{7'}$: $(\delta\Cc{7},\delta\Cc{7'})=\Cc{7}^{\rm SM}(0.2 e^{i\theta},0.3e^{i\theta})$\,.
\end{itemize}
These scenarios are chosen for illustration purpose, but the size of the NP effects considered is compatible with the results obtained from dedicated experimental~\cite{LHCb:2020dof,LHCb:2021byf} and theoretical~\cite{Paul:2016urs} studies, as well as from recent global fits ~\cite{Alguero:2021anc, Altmannshofer:2021qrr, Hurth:2021nsi, Geng:2021nhg, Ciuchini:2020gvn}.

\begin{figure}[!t]
\begin{center}
\includegraphics[width=0.4\textwidth]{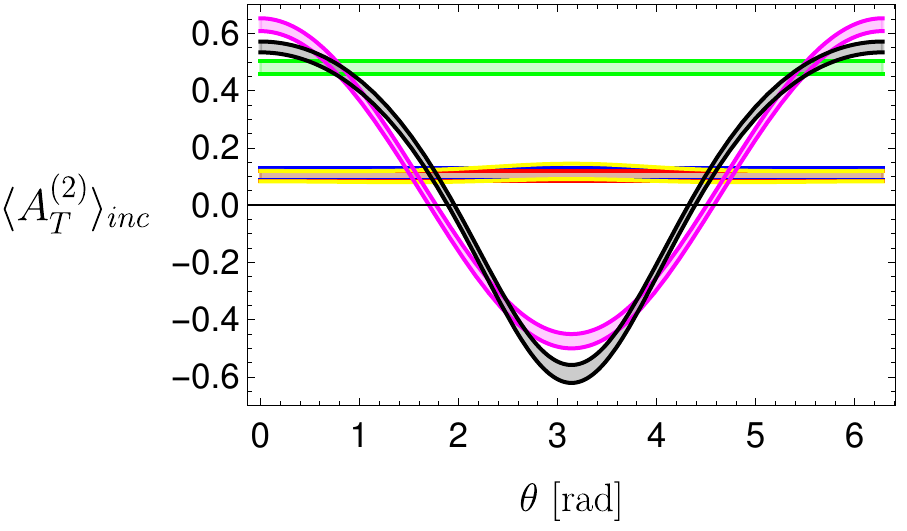} ~\includegraphics[width=0.58\textwidth]{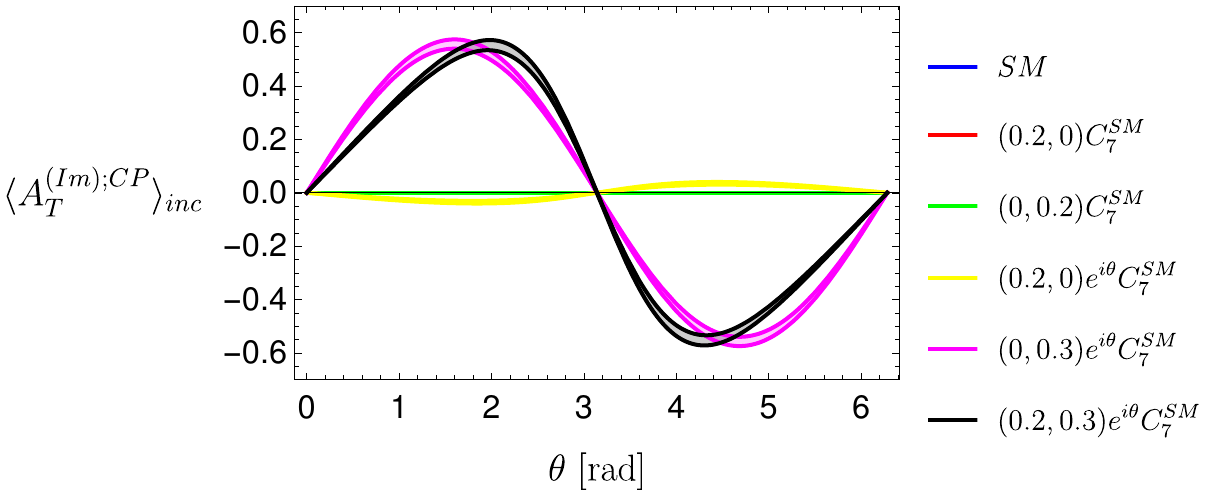}
\end{center}
\caption{\sl\small Variation with the NP weak phase $\theta$ of $\langle A_T^{(2)}\rangle_{\rm inc} $ and $\langle A_T^{(\text{Im});CP}\rangle_{\rm inc}$ in the bin $q^2\in [0.0008,0.257]$ GeV$^2$. The selected NP scenarions are shown, taking into account meson-mixing effects. Note, in particular, that the second and third scenarios have no impact on $\langle A_T^{(\text{Im});CP}\rangle_{\rm inc}$.
\label{fig:Transverse-Impact}
}
\end{figure}

\begin{figure}[!t]
\begin{center}
{\includegraphics[width=0.4\textwidth]{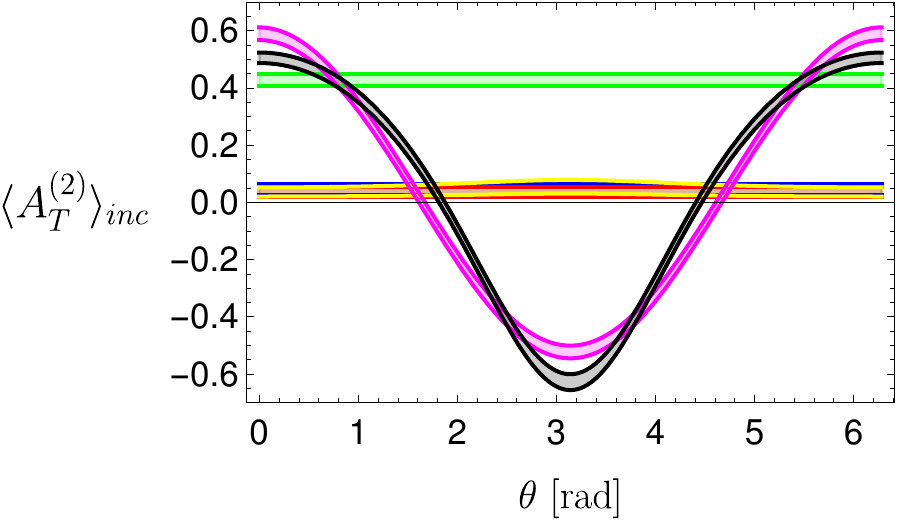}
\put(-135,35) {\footnotesize\color{gray} no mixing}}
~{\includegraphics[width=0.58\textwidth]{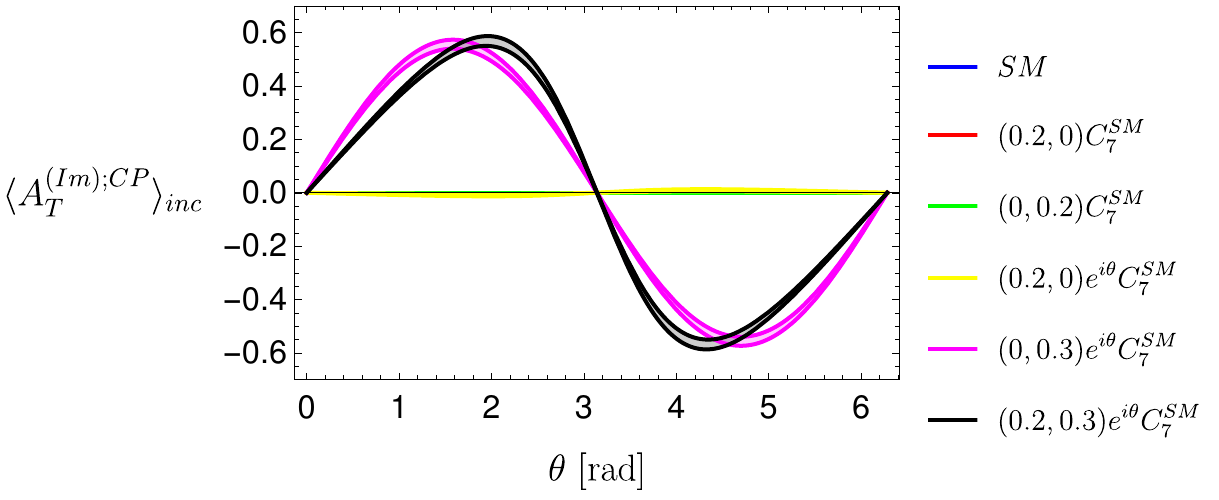}
\put(-205,35) {\footnotesize\color{gray} no mixing}}
\end{center}
\caption{\sl\small Variation with the NP weak phase of $\langle A_T^{(2)}\rangle_{\rm inc} $ and $\langle A_T^{(\text{Im});CP}\rangle_{\rm inc}$ in the bin $q^2\in [0.0008,0.257]$ GeV$^2$. For comparison, the selected NP scenarios are shown without taking into account meson-mixing effects.
\label{fig:Transverse-Impact-no-mixing}
}
\end{figure}

In Sec.~\ref{sec:nearSM}, we argued that
 $\smash{\langle A_T^{(2)}\rangle_{\rm inc}}$ and $\smash{\langle A_T^{(\text{Im});CP}\rangle_{\rm inc}}$ are the main observables of interest for the scenarios defined above. In Fig.~\ref{fig:Transverse-Impact}, we show how these quantities, binned over $q^2$ between 0.0008 and 0.257 GeV$^2$, vary as a function of the NP phase $\theta$.  For comparison, the same plots are shown in Fig.~\ref{fig:Transverse-Impact-no-mixing} without taking the mixing effects into account.
We see that 
$\smash{A_T^{(2)}}$ and $\smash{A_T^{(\text{Im});CP}}$ exhibit a significant sensitivity to NP phases under the various scenarios considered. Moreover, the neutral-meson mixing affects noticeably $A_T^{(2)}$, leading to a change of the central value from 0.04 to 0.10 in the SM, whereas $A_T^{(\text{Im});CP}$ is only marginally affected. Finally, we mention again that other transverse asymmetries can be obtained from $d\Gamma-d\bar\Gamma$ and they are discussed in Appendix~\ref{app:asymmbis}, exhibiting a very large sensitivity to neutral-meson mixing.

\section{Conclusion}\label{sec:conclusion}

In the context of $b\to s\ell\ell$ anomalies, it is particularly important to analyse the related $b\to s\gamma$ decay
which are expected to be  particularly sensitive to New Physics contributions modifying the short-distance Wilson coefficients $\Cc{7}$ and $\Cc{7'}$, describing this decay in the Low-Energy Effective Field Theory at the scale of the $b$-quark mass. Among several decays modes and approaches, one can use $B_s\to\phi ee$ at low $q^2$, in a kinematic range where the photon pole dominates. In this range, the longitudinal amplitude gets suppressed as well as many angular observables, compared to the angular observables involving only transverse amplitudes (i.e.,~$J_i$ with $i=1s,2s,3,9$).

The presence of $B_s-\bar{B}_s$ mixing affects the theoretical description of the angular observables that can be determined from this mode. A time-dependent modulation arises and involves interference terms between mixing and decay which are sensitive to the relative moduli and phase of $\Cc{7}$ and $\Cc{7'}$. Therefore, these interference terms open an interesting window to constrain not only the size of the Wilson coefficients, but also their (weak) phases. We exploited the results of Ref.~\cite{Descotes-Genon:2015hea} in the photon-pole dominance approximation to determine simplified expressions of these interference terms, associated with the time dependence of the angular coefficients $J_{1s}$, $J_{2s}$, $J_{3}$ and $J_{9}$.  In the photon-pole approximation, the time-dependent terms associated with $J_{1s}$ and $J_{2s}$ contain the same information as the CP asymmetries that can be built for $B_s\to\phi\gamma$, whereas the other angular observables feature mixing-induced time-dependence that can be exploited to probe complementary combinations of Wilson coefficients.

These interference terms can be extracted through a time-dependent analysis at Belle II. We also demonstrated that they affect time-integrated observables at LHCb and, more generally, at any collider producing $b\bar{b}$ pairs incoherently. Within the approximation where the photon-pole dominates and assuming that NP does not alter the overall hierarchy between $\Cc{7}$ and $\Cc{7'}$, we showed that the time-integrated value of $A_T^{(2)}$ and
$A_T^{(\text{Im});CP}$
receive terms of similar size from mixing-independent and mixing-induced terms, whereas the other usual transverse asymmetries are not significantly affected. Moreover, we discussed briefly the three time-integrated transverse asymmetries that can be measured in $d\Gamma-d\bar\Gamma$, and we pointed out that two of them, corresponding to $A_T^{(2);CP}$ and $A_T^{(\text{Im})}$, 
are dominated by neutral-meson mixing at low $q^2$ and thus require a specific interpretation in terms of the Wilson coefficients of the low-energy effective field theory.

Finally, we have considered several NP scenarios involving weak phases for $\Cc{7}$ and/or $\Cc{7'}$. We showed that $A_T^{(2)}$  and $\smash{A_T^{(\text{Im});CP}}$ measured at low $q^2$ could indeed provide interesting constraints, and that it was important to take into account mixing-induced contributions for the foreseen time-integrated measurements at LHCb.
In particular, mixing effects are noticeable for $\smash{A_T^{(2)}}$ already within the SM. 
The $\smash{A_T^{(\text{Re});CP}}$ transverse asymmetry has not been considered in this discussion, as it is mainly sensitive weak phases in $\Cc{10^{(')}}$. Even though we have not discussed this third asymmetry in detail given our choices of NP scenarios, it is worth stressing that such a measurement would provide complementary information to  $\smash{A_T^{(2)}}$ and $\smash{A_T^{(\text{Im});CP}}$.

Given that LHCb has already performed angular analyses of $B_s\to \phi\mu\mu$, as well as $B\to K^*ee$ at low $q^2$, one may hope that the angular analysis of $B_s\to\phi ee$ at low $q^2$ will be soon available in a time-integrated form, requiring
to include the effects of neutral-meson mixing for their theoretical interpretation. On a longer term, Belle II could measure directly the time dependence of this decay to extract the modulation terms $s_i$ and $h_i$, which are not affected by strong phases. Both types of observables  could provide very useful additional information on $b\to s\gamma$ decays, helping us to determine if they are significantly affected by New Physics contributions and if they can contribute to the ongoing debate on the nature of the $b$-quark anomalies.

\section*{Acknowledgements}

We thank Marie-Hélène Schune and Gaëlle Khreich for valuable comments and discussions. We also thank Peter Stangl for useful exchanges about {\tt flavio}~\cite{Straub:2018kue}. This project has received support from the European Union’s Horizon 2020 research and innovation programme under the Marie Sklodowska-Curie grant agreement No 860881-HIDDeN. I.~P.~receives funding from ``P2IO LabEx (ANR-10-LABX-0038)"  in the framework ``Investissements d'Avenir" (ANR-11-IDEX-0003-01) managed by the Agence Nationale de la Recherche (ANR), France

\clearpage

\appendix

\section{Angular conventions}
\label{app:conventions}

\begin{figure}[!t]
\begin{center}
\includegraphics[scale=0.75]{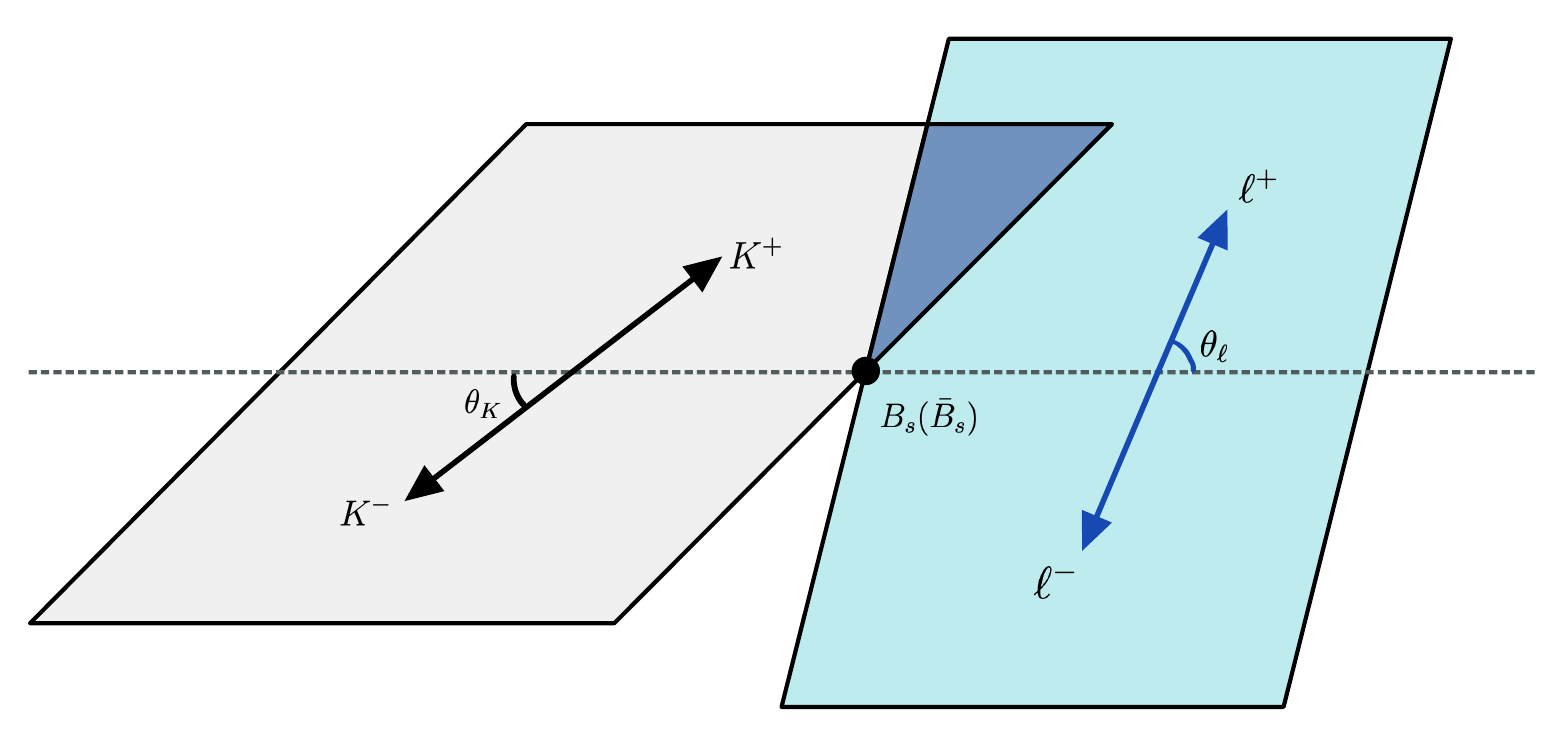}
\end{center}
\caption{\sl\small
Angular convention for the decays $B_{s} \to \phi (\to K^- K^+)\ell\ell$ and $\bar{B}_{s} \to \phi (\to K^- K^+)\ell\ell$ . The angles $\theta_\ell$ and $\theta_K$ are defined in the di-lepton and $\phi$ rest-frames, respectively. The angle $\phi$ is defined between the $K^+ K^-$ and $\ell^+\ell^-$ in the $B$ rest frame and, although not shown in this diagram, it can be unambiguously extracted from Eq.~\eqref{eq:vec-leptons} and \eqref{eq:vec-mesons}.
\label{fig:angular}
}
\end{figure}

Our angular conventions for the decays $B_{s} \to \phi (\to K^+ K^-)\ell\ell$ and $\bar{B}_{s} \to \phi (\to K^+ K^-)\ell\ell$ are summarized in Fig.~\ref{fig:angular} adapting the angular conventions from Ref.~\cite{Gratrex:2015hna,Becirevic:2016zri}. The leptonic and hadronic four-vectors are defined in the $B_s$ $(\overline{B}_s)$ rest frames as $q^\mu = (q_0,0,0,q_z)$ and $k^\mu = (k_0,0,0,-q_z)$, respectively, with
\begin{align}
q_0 = \dfrac{M^2+q^2-m^2}{2M}\,,\qquad\quad k_0 = \dfrac{M^2-q^2+m^2}{2M}\,,\qquad\quad q_z = \dfrac{\lambda^{1/2}(M^2,m^2,q^2)}{2 M}\,,
\end{align}
where $\lambda(a^2,b^2,c^2)=(a^2-(b-c)^2)(a^2-(b+c)^2)$, $M$ denotes the $B_s$-meson mass and $m$ is the $\phi$ mass. In the dilepton rest frame, the leptonic four-vectors read
\begin{align}
\begin{split}
q_1^\mu &=(E_\ell, +|p_\ell|\sin \theta_\ell \cos\phi, +|p_\ell|\sin \theta_\ell \sin\phi, -|p_\ell| \cos\theta_\ell)\,,\\[0.3em]
q_2^\mu &=(E_\ell, -|p_\ell|\sin \theta_\ell \cos\phi, -|p_\ell|\sin \theta_\ell \sin\phi, +|p_\ell| \cos\theta_\ell)\,,
\label{eq:vec-leptons}
\end{split}
\end{align}
where $q_1(q_2)$ denotes the momentum of the negative (positive) charge lepton, $E_\ell=\sqrt{q^2}/2$ and $|p_\ell|=\sqrt{q^2}/2\,\beta_\ell$, with $\beta_\ell=\sqrt{1-4 m_{\ell}^2/q^2}$. Similarly, the hadronic vectors can be written in the $\phi$-meson rest frame as
\begin{align}
\begin{split}
k_1^\mu &= (E_k, +|p_K|\sin \theta_K, 0, -|p_K| \cos\theta_K)\,,\\[0.3em]
k_2^\mu &= (E_k, -|p_K|\sin \theta_K, 0, +|p_K| \cos\theta_K)\,,
\label{eq:vec-mesons}
\end{split}
\end{align}
where $k_1(k_2)$ stands for the momentum of $K^-(K^+)$, and $E_K=m/2$ and $|p_K|=m/2\,\sqrt{1-4 m_K^2/m^2}$ for an on-shell $\phi$-meson.

\section{Transversity amplitudes}
\label{app:transversity}

In this Appendix, we provide for completeness our expressions for the transversity amplitudes for the $B_s\to \phi\ell\ell$ decays~\cite{Altmannshofer:2008dz},
\begin{align}
A_{\perp}^{L(R)} &= {\cal N}_{V} \sqrt{2} \lambda_V^{1/2}\left[[(C_9+C_9')\mp(C_{10}+C_{10}')]\frac{V(q^2)}{M+m}+\frac{2 m_b}{q^2}(C_7+C_7') T_1(q^2)\right]\,, \\[0.45em]
A_{\parallel}^{L(R)} &=  -{\cal N}_{V} \sqrt{2}(M^2-m^2)\left[[(C_9-C_9')\mp (C_{10}-C_{10}')]\frac{A_1(q^2)}{M-m}+\frac{2 m_b}{q^2}(C_7-C_7')T_2(q^2)\right]\,,\\[0.5em]
A_0^{L(R)}&=-\frac{{\cal N}_{V}}{2 m \sqrt{q^2}}\Big{\lbrace} 2 m_b (C_7-C_7')\left[(M^2+3m^2-q^2)T_2(q^2)-\frac{ \lambda_V T_3(q^2)}{M^2-m^2} \right] \\
&+[(C_9-C_9')\mp(C_{10}-C_{10}')]\cdot\left[ (M^2-m^2-q^2)(M+m)A_1(q^2)-\frac{ \lambda_V A_2(q^2)}{M+m}\right] \Big{\rbrace}\,, \nonumber\\[0.45em]
A_{t} &= \dfrac{\mathcal{N}_V}{\sqrt{q^2}}\lambda_V^{1/2}\left[2 (C_{10}-C_{10}^\prime)+\dfrac{q^2}{m_\ell}(C_P-C_{P^\prime})\right] A_0(q^2)\,,\\[0.45em]
A_{S} &= -2 \mathcal{N}_V \lambda_V^{1/2} (C_S-C_{S^\prime}) A_0(q^2)\,,
\end{align}
where $M$ and $m$ denote the $B_s$ and $\phi$ masses, respectively, and the normalization  $\mathcal{N}_V$ is given by
\begin{equation}
{\cal N}_{V}=|V_{tb}V_{ts}^\ast|\left[ \frac{ G_F^2 \alpha^2}{3 \times 2^{10} \pi^5 M^3} \lambda_V^{1/2} q^2 \beta_\ell \right]^{1/2}.
\end{equation}
where we write $\lambda_V=\lambda(M,m,\sqrt{q^2})$, with $\lambda(a,b,c)=[a^2-(b-c)^2] [a^2-(b+c)^2]$. Furthermore, we consider the usual conventions for the $B\to V$ form-factors~\cite{Altmannshofer:2008dz}.

\section{Angular coefficients in the photon-pole approximation}
\label{app:photon-pole}

In this Appendix, we provide the expressions for the non-vanishing angular observables described in Sec.~\ref{sec:timeintobs} in the photon-pole approximation:

\begin{eqnarray}
    \langle S_{1s}\rangle_{\rm inc} &\sim& \frac{3}{4} \frac{\alpha_{-1/3}}{\alpha_{-1/2}} \,,\\[0.4em]
    \langle S_{2s}\rangle_{\rm inc} &\sim& \frac{1}{4} \frac{\alpha_{1}}{\alpha_{-1/2}}\,,\\[0.4em] 
  \langle S_3\rangle_{\rm inc} &\sim&
   \frac{1}{2}  \frac{\alpha_{1}}{\alpha_{-1/2}}\frac{[|C_+|^2+|\bar{C}_+|^2 - |C_-|^2- |\bar{C}_-|^2]+y\cdot 2 \text{Re}[\bar{C}_+C_+^*+\bar{C}_-C_-^*]}
    {[|C_+|^2 +|\bar{C}_+|^2 + |C_-|^2 +|\bar{C}_-|^2 ]+ y\cdot 2 \text{Re}[{\bar{C}_+C_+^*-\bar{C}_-C_-^*]}}\,,\\[0.4em]
   \langle  S_9\rangle_{\rm inc} &\sim&  \frac{\alpha_{1}}{\alpha_{-1/2}} \frac{1-y^2}{1+x^2}
      \frac{-\text{Im}[C_-^* C_+ + \bar{C}_-^* \bar{C}_+] +x\cdot  \text{Re}[\bar{C}_-C_+^*+\bar{C}_+^*C_-]}
      {[|C_+|^2 +|\bar{C}_+|^2 + |C_-|^2 +|\bar{C}_-|^2 ]+ y\cdot 2 \text{Re}[{\bar{C}_+C_+^*-\bar{C}_-C_-^*]}}
     \,,
\end{eqnarray}
and
\begin{eqnarray}
   \langle A_{1s}\rangle_{\rm inc} &\sim&  \frac{3}{4}  \frac{\alpha_{-1/3}}{\alpha_{-1/2}} 
    \frac{1-y^2}{1+x^2}
      \frac{[|C_+|^2 -|\bar{C}_+|^2 + |C_-|^2 -|\bar{C}_-|^2 ]+x\cdot 2\text{Im}[\bar{C}_+C_+^*-\bar{C}_-C_-^*]}
      {[|C_+|^2 +|\bar{C}_+|^2 + |C_-|^2 +|\bar{C}_-|^2 ]+ y\cdot 2 \text{Re}[{\bar{C}_+C_+^*-\bar{C}_-C_-^*]}}\,,\\[0.4em]
   \langle  A_{2s}\rangle_{\rm inc} &\sim&  \frac{1}{4}   \frac{\alpha_{1}}{\alpha_{-1/2}}
    \frac{1-y^2}{1+x^2}
      \frac{[|C_+|^2 -|\bar{C}_+|^2 + |C_-|^2 -|\bar{C}_-|^2 ]+x\cdot 2\text{Im}[\bar{C}_+C_+^*-\bar{C}_-C_-^*]}
      {[|C_+|^2 +|\bar{C}_+|^2 + |C_-|^2 +|\bar{C}_-|^2 ]+ y\cdot 2 \text{Re}[{\bar{C}_+C_+^*-\bar{C}_-C_-^*]}}\,,\\[0.4em]
   \langle  A_3\rangle_{\rm inc} &\sim&  \frac{1}{2}  \frac{\alpha_{1}}{\alpha_{-1/2}}
    \frac{1-y^2}{1+x^2}
      \frac{[|C_+|^2 -|\bar{C}_+|^2 - |C_-|^2 + |\bar{C}_-|^2 ]+x\cdot 2\text{Im}[\bar{C}_+C_+^*+\bar{C}_-C_-^*]}
      {[|C_+|^2 +|\bar{C}_+|^2 + |C_-|^2 +|\bar{C}_-|^2 ]+ y\cdot 2 \text{Re}[{\bar{C}_+C_+^*-\bar{C}_-C_-^*]}}\,,\\[0.4em]
 \langle A_9\rangle_{\rm inc} &\sim&  \frac{\alpha_{1}}{\alpha_{-1/2}} \frac{-\text{Im}[C_-^* C_+ - \bar{C}_-^* \bar{C}_+]  - y\cdot \text{Im}[\bar{C}_-C_+^*-\bar{C}_+^*C_-]}
    {[|C_+|^2 +|\bar{C}_+|^2 + |C_-|^2 +|\bar{C}_-|^2 ]+ y\cdot 2 \text{Re}[{\bar{C}_+C_+^*-\bar{C}_-C_-^*]}}  \,.
\end{eqnarray}

\section{Transverse asymmetries built from $d\Gamma-d\bar\Gamma$} 
\label{app:asymmbis}

In this Section, we discuss the following asymmetries which are available if one had access to  $d\Gamma-d\bar\Gamma$.
\begin{equation}
 A_{T}^{(2);CP}=\frac{A_{3}}{2S_{2s}}\,, \qquad A_T^{(\text{Re})}=\frac{S_{6s}}{4S_{2s}}\,, \qquad
 A_{T}^{(\text{Im})}=\frac{S_{9}}{2S_{2s}}\,.
\end{equation}
We provide expressions for these asymmetries under the various hypotheses considered in the main text, and we explore their sensitivity to NP in $\Cc{7}$ and $\Cc{7'}$. Their expression in terms of the transversity amplitudes is given by
\begin{eqnarray}
A_{T}^{(2);CP}&=&\frac{|A^L_\perp|^2+|A^R_\perp|^2-|A_{||}^L|^2-|A_{||}^R|^2-(A \leftrightarrow \bar{A})}{|A^L_\perp|^2+|A^R_\perp|^2+|A_{||}^L|^2+|A_{||}^R|^2+(A \leftrightarrow \bar{A})}\,,\\[0.3em]
A_T^{(\text{Re})}&=&\frac{2{\rm Re}[(A_{||}^LA_\perp^{L*}-A_{||}^R A_\perp^{R*})+(A \leftrightarrow \bar{A})]}{|A^L_\perp|^2+|A^R_\perp|^2+|A_{||}^L|^2+|A_{||}^R|^2+(A \leftrightarrow \bar{A})}\,,\\[0.3em]
A_T^{(\text{Im})}&=&-\frac{2{\rm Im}\big{[}(A_{||}^LA_\perp^{L*}+A_{||}^R A_\perp^{R*})+(A \leftrightarrow \bar{A})\big{]}}{|A^L_\perp|^2+|A^R_\perp|^2+|A_{||}^L|^2+|A_{||}^R|^2+(A \leftrightarrow \bar{A})}\,.
\end{eqnarray}

\noindent Using the decomposition given in Eq.~(\ref{eq:ampldecomposition}), we obtain
\begin{eqnarray}
A_{T}^{(2);CP}&=&
\frac{2}{D}\Big[
2h^\perp\rho^\perp_{79}\sin\delta^\perp \sin\phi^\perp_{79}-(\perp \leftrightarrow ||)
\Big]\,,\\[0.4em]
A_T^{(\text{Re})}&=&-\frac{2}{D}\Big[
\rho^\perp_{79}\rho^{||}_{10}\cos(\phi^{\perp}_{79}-\phi^{||}_{10})+h^\perp\rho^{||}_{10}\cos\delta^\perp\cos\phi^{||}_{10}+(\perp \leftrightarrow ||) \Big{]}\\[0.4em]
A_T^{(\text{Im})}&=&
\frac{1}{D}\Big[
h^\perp h^{||} \sin(\delta^\perp-\delta^{||})
+2 h^\perp \rho^{||}_{79} \sin\delta^\perp\cos\phi^{||}_{79}
-(\perp \leftrightarrow ||) 
\Big]\,,
\end{eqnarray}
with the denominator $D$ defined in Eq.~\eqref{eq:Dparameter}. From these expressions,
we see that $A_T^{(\text{Re})}$ is proportional to the contribution from $\Cc{10^{(')}}$ and will thus not be relevant in the photon-pole approximation. Indeed, in this approximation, the asymmetries become
\begin{eqnarray}\label{eq:AT2CPgeneral}
     \langle A_T^{(2);CP}\rangle_{\rm inc} &\sim& 
     \frac{1-y^2}{1+x^2}
      \frac{[|C_+|^2 -|\bar{C}_+|^2 - |C_-|^2 + |\bar{C}_-|^2 ]+x\cdot 2\text{Im}[\bar{C}_+C_+^*+\bar{C}_-C_-^*]}
      {[|C_+|^2 +|\bar{C}_+|^2 + |C_-|^2 +|\bar{C}_-|^2 ]+ y\cdot 2 \text{Re}[{\bar{C}_+C_+^*-\bar{C}_-C_-^*]}}\,,\\[0.3em]
     \langle A_T^{(\text{Re})}\rangle_{\rm inc}  &\sim & 0 \,,\\[0.3em]
     \langle A_T^{(\text{Im})}\rangle_{\rm inc}  &\sim & 
      2 \frac{1-y^2}{1+x^2}
      \frac{-\text{Im}[C_-^* C_+ + \bar{C}_-^* \bar{C}_+] +x\cdot  \text{Re}[\bar{C}_-C_+^*+\bar{C}_+^*C_-]}
      {[|C_+|^2 +|\bar{C}_+|^2 + |C_-|^2 +|\bar{C}_-|^2 ]+ y\cdot 2 \text{Re}[{\bar{C}_+C_+^*-\bar{C}_-C_-^*]}}
      \,,\label{eq:ATImgeneral}
\end{eqnarray}
and near the SM point, they reduce further down to
\begin{eqnarray}\label{eq:AT2CPnearSM}
\langle A_T^{(2);CP}\rangle_{\rm inc}&\sim&  
     \frac{2}{x}\frac{\text{Re}[\Cc{7}]\text{Im}[\Cc{7}]}{(\text{Re}[\Cc{7}])^2+(\text{Im}[\Cc{7}])^2}\,,\\[0.3em]
\langle A_T^{(\text{Re})}\rangle_{\rm inc}   &\sim & 0\,,\\[0.3em]
\langle A_T^{(\text{Im})}\rangle_{\rm inc}
&\sim&
\frac{1}{x}\frac{(\text{Re}[\Cc{7}])^2-(\text{Im}[\Cc{7}])^2}{(\text{Re}[\Cc{7}])^2+(\text{Im}[\Cc{7}])^2}\,.\label{eq:ATimnearSM}
\end{eqnarray}
As already discussed in Sec.~\ref{sec:nearSM}, these expressions match the mixing-induced contribution to $\langle A_T^{(2)}\rangle_{\rm inc}$ and $\langle A_T^{(\text{Im});CP}\rangle_{\rm inc}$ (up to a factor $1/x$ instead of $y$). On one hand, the measurement through Eqs.~(\ref{eq:AT2CPnearSM}) and (\ref{eq:ATimnearSM}) provides directly this contribution whereas it must be disentangled from the mixing-independent term in Eqs.~(\ref{eq:AT2nearSM}) and (\ref{eq:ATimCPnearSM}). On the other hand, Eqs.~(\ref{eq:AT2CPnearSM}) and (\ref{eq:ATimnearSM}) require flavour tagging in order to build $d\Gamma-d\bar\Gamma$ which makes this measurement much more challenging than Eqs.~(\ref{eq:AT2nearSM}) and (\ref{eq:ATimCPnearSM}).

\begin{figure}[!t]
\begin{center}
\includegraphics[width=0.42\textwidth]{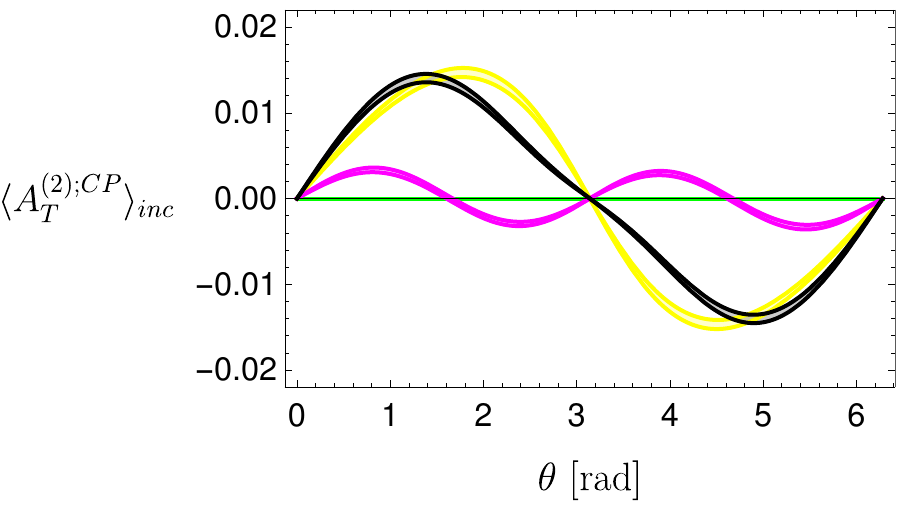} ~\includegraphics[width=0.55\textwidth]{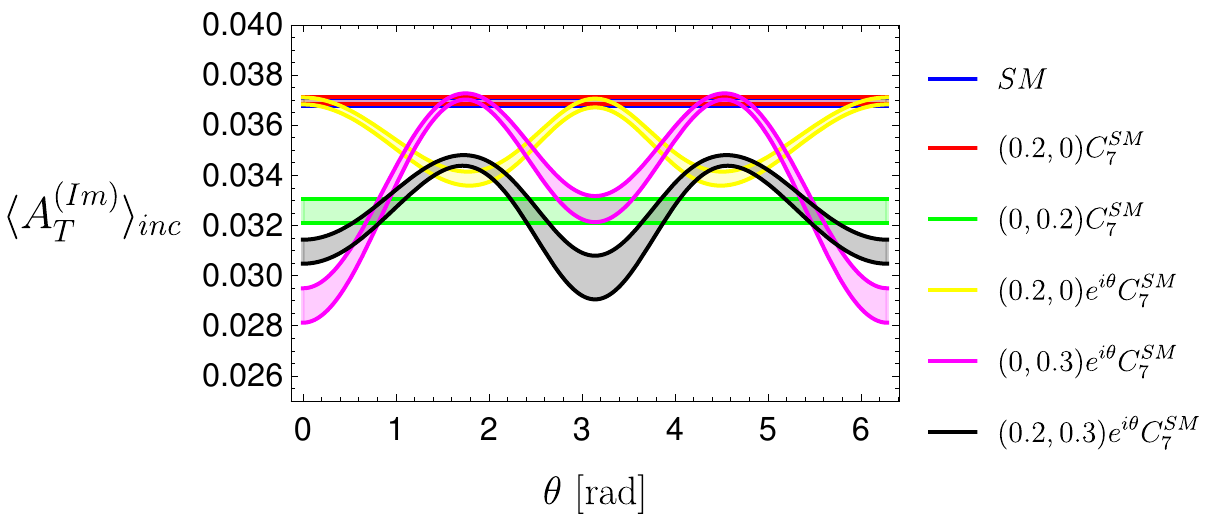}
\end{center}
\caption{\sl\small Variation with $\theta$ of $\langle A_T^{(2);CP}\rangle_{\rm inc} $ and $\langle A_T^{(\text{Im})}\rangle_{\rm inc}$ in the bin $q^2\in [0.0008,0.257]$ GeV$^2$. The selected NP scenarios are shown taking into account meson-mixing effects.
\label{fig:Transverse-Impact-2}
}
\end{figure}

\begin{figure}[!t]
\begin{center}
{\includegraphics[width=0.42\textwidth]{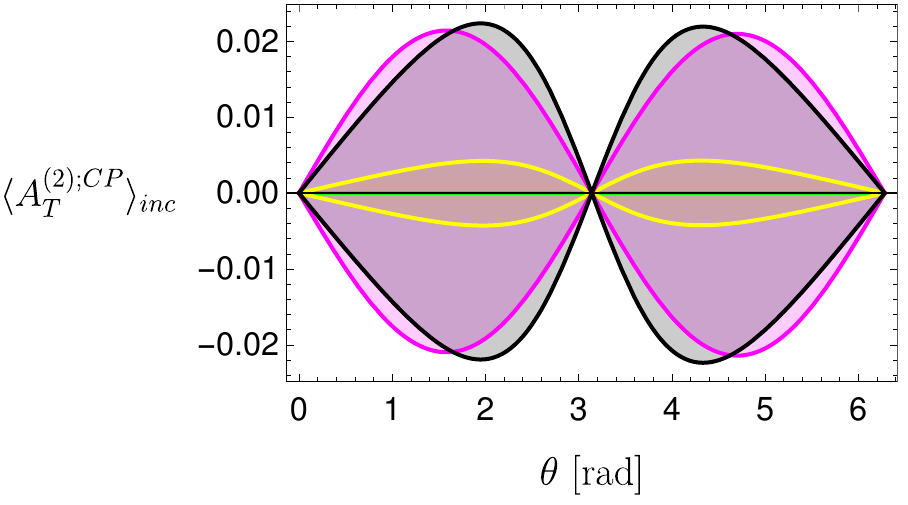}
\put(-135,35) {\footnotesize\color{gray} no mixing}}~
{\includegraphics[width=0.55\textwidth]{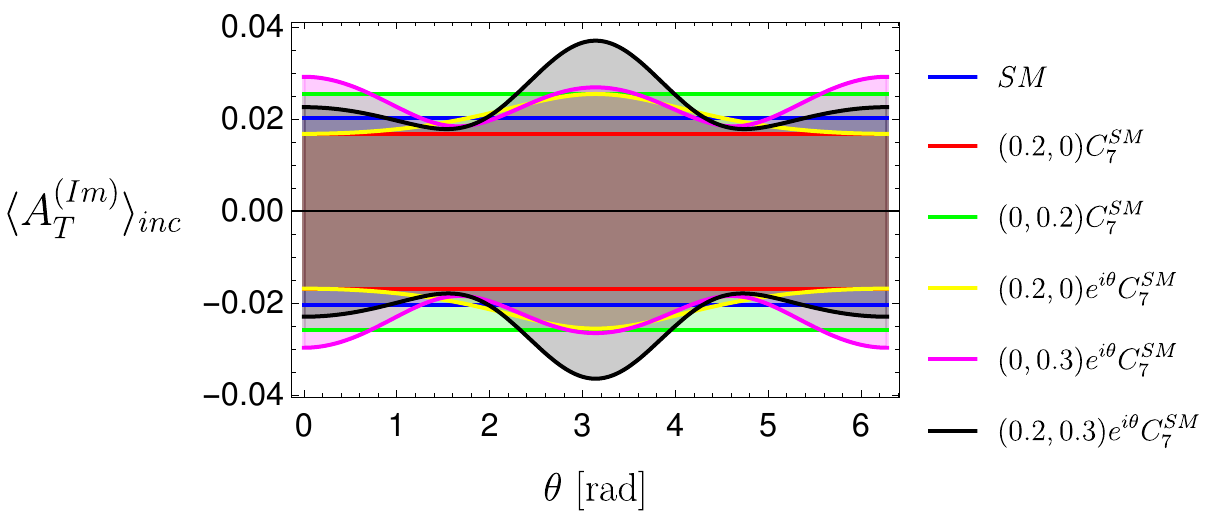}
\put(-195,35) {\footnotesize\color{gray} no mixing}}
\end{center}
\caption{\sl\small Variation with $\theta$ of $\langle A_T^{(2);CP}\rangle_{\rm inc} $ and $\langle A_T^{(\text{Im})}\rangle_{\rm inc}$ in the bin $q^2\in [0.0008,0.257]$ GeV$^2$. For comparison, the selected NP scenarios are shown without taking into account meson-mixing effects.
\label{fig:Transverse-Impact-no-mixing-2}
}
\end{figure}

We illustrate the sensitivity to a NP phase in $\Cc{7}$ and $\Cc{7'}$ for these quantities in Figs.~\ref{fig:Transverse-Impact-2} (with mixing) and \ref{fig:Transverse-Impact-no-mixing-2} (without mixing).
The mixing-independent part shown in Fig.~\ref{fig:Transverse-Impact-no-mixing-2} and obtained by taking $x,y\to 0$ in Eqs.~(\ref{eq:AT2CPgeneral})-(\ref{eq:ATImgeneral}) becomes actually $1/x^2$ suppressed once mixing is taken into account, so that
 neutral-meson mixing dominates completely these observables and leads to a very different pattern shown in Fig.~\ref{fig:Transverse-Impact-2}. This strikingly different behaviour highlights the importance of taking into account neutral-meson mixing to constrain Wilson coefficients using these two asymmetries.

We do not consider $\langle A_T^{(\text{Re})}\rangle_{\rm inc}$ as we discussed only scenarios with NP in $\Cc{7}$ and $\Cc{7'}$ and focused on quantities that are mainly sensitive to these coefficients in the photon-pole approximation. Naturally,  $\langle A_T^{(\text{Re})}\rangle_{\rm inc}$ is important for scenarios involving NP in a larger set of Wilson coefficients, as illustrated for instance  in Ref.~\cite{Becirevic:2011bp} in the case of the self-tagging mode $B\to K^*ee$ where neutral-meson mixing does not play a role.

\end{document}